\documentclass[sigconf]{acmart}
\usepackage{graphicx}
\usepackage{subcaption}
\usepackage{float}
\usepackage{multirow}
\usepackage{array}
\usepackage{makecell}
\usepackage{xcolor}
\definecolor{brownishred}{RGB}{0,0,0}
\newcommand{\colorchange}[1]{{\textcolor{brownishred}{#1}}}

\AtBeginDocument{%
  }

\setcopyright{acmlicensed}
\copyrightyear{2025}
\acmYear{2025}
\setcopyright{cc}
\setcctype{by-nd}
\acmConference[CHI '25]{CHI Conference on Human Factors in Computing Systems}{April 26-May 1, 2025}{Yokohama, Japan}
\acmBooktitle{CHI Conference on Human Factors in Computing Systems (CHI '25), April 26-May 1, 2025, Yokohama, Japan}\acmDOI{10.1145/3706598.3713847}
\acmISBN{979-8-4007-1394-1/25/04}

\begin{document}

\title[VisiMark]{VisiMark: Characterizing and Augmenting Landmarks for People with Low Vision in Augmented Reality to Support Indoor Navigation}

\author{Ruijia Chen}
\affiliation{%
  \institution{University of Wisconsin-Madison}
  \city{Madison}
  \state{Wisconsin}
  \country{USA}
}
\email{ruijia.chen@wisc.edu}
\orcid{0000-0002-1655-6228}

\author{Junru Jiang}
\affiliation{%
  \institution{University of Wisconsin-Madison}
  \city{Madison}
  \state{Wisconsin}
  \country{USA}
  }
\email{jjiang324@wisc.edu}

\author{Pragati Maheshwary}
\affiliation{%
  \institution{Carnegie Mellon University}
  \city{Pittsburgh}
  \state{Pennsylvania}
  \country{USA}
}
\email{pragati2@andrew.cmu.edu}

\author{Brianna R. Cochran}
\affiliation{%
  \institution{University of Wisconsin-Madison}
  \city{Madison}
  \state{Wisconsin}
  \country{USA}
  }
\email{bcochran2@wisc.edu}

\author{Yuhang Zhao}
\affiliation{%
  \institution{University of Wisconsin-Madison}
  \city{Madison}
  \state{Wisconsin}
  \country{USA}
  }
\email{yuhang.zhao@cs.wisc.edu}

\renewcommand{\shortauthors}{Chen et al.}

\begin{abstract}
Landmarks are critical in navigation, supporting self-orientation and mental model development. Similar to sighted people, people with low vision (PLV) frequently look for landmarks via visual cues but face difficulties identifying some important landmarks due to vision loss. We first conducted a formative study with six PLV to characterize their challenges and strategies in landmark selection, identifying their unique landmark categories (e.g., area silhouettes, accessibility-related objects) and preferred landmark augmentations. We then designed \textit{VisiMark}, an AR interface that supports landmark perception for PLV by providing both overviews of space structures and in-situ landmark augmentations. We evaluated VisiMark with 16 PLV and found that VisiMark enabled PLV to perceive landmarks they preferred but could not easily perceive before, and changed PLV's landmark selection from only visually-salient objects to cognitive landmarks that are more important and meaningful. We further derive design considerations for AR-based landmark augmentation systems for PLV.
\end{abstract}


%TC:ignore
\begin{teaserfigure}
  \includegraphics[width=\textwidth]{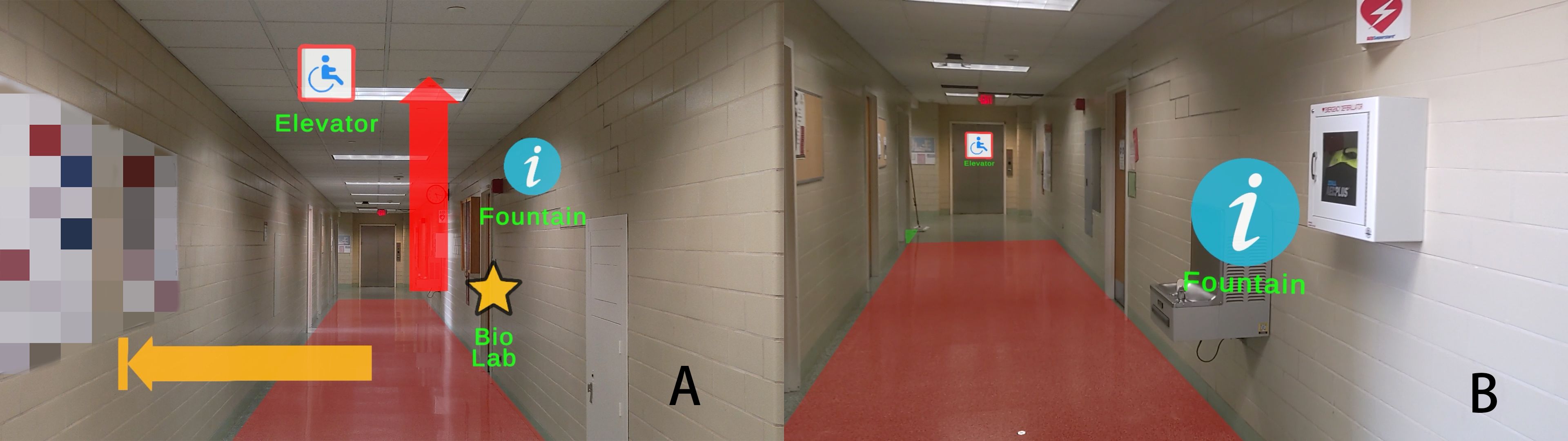}
    % \vspace{-5ex}
  \caption{\textit{VisiMark} provides landmark augmentations on head-mounted AR to support wayfinding and mental map construction. VisiMark includes two features: (A) \textit{Signboard}, an overview of hallway structures and upcoming landmarks at intersections, and (B) \textit{In-situ Labels}, world-anchored icons and texts to highlight the types and positions of landmarks in the physical environment.}
  \Description{This image shows VisiMark system. VisiMark provides landmark augmentations on head-mounted AR to support wayfinding and mental map construction. VisiMark includes two features: (A) Signboard, an overview of hallway structures and upcoming landmarks at intersections, and (B) In-situ Labels, world-anchored icons and texts to highlight the types and positions of landmarks in the physical environment.}
  \label{fig:teaser}
\end{teaserfigure}
%TC:endignore


\maketitle

\section{INTRODUCTION}

Low vision refers to visual impairments that fall short of blindness but cannot be corrected with eyeglasses or contacts, affecting the independence and quality of life of 596 million people worldwide \cite{IAPBVisionAtlas}. Unlike blind people, people with low vision (PLV) rely on their functional vision in daily activities \cite{szpiro2016finding}. However, they experience visual challenges due to various low vision conditions, such as loss of central or peripheral vision, blurred vision, and reduced contrast sensitivity \cite{AOALowVision}. These conditions significantly impact daily tasks, such as reading \cite{wang2023understanding, wang2024gazeprompt}, navigating \cite{szpiro2016finding, zhao2020effectiveness}, housework \cite{jones2019analysis, wang2023practices}, and socializing \cite{senra2015psychologic, shah2020association}.

Navigation is a complex and challenging task for PLV \cite{corn2010foundations,szpiro2016finding}, and landmarks play a critical role in this process \cite{siegel1975development}. Landmarks are commonly defined as stationary, distinct, and salient objects or places in the surrounding space that are more likely to be selected as spatial references \cite{yesiltepe2021landmarks, millonig2007developing}. They enable self-orientation \cite{raubal2002enriching}, convey spatial information \cite{tom2003referring,lovelace1999elements}, and support mental model development \cite{michon2001and, millonig2007developing}. Prior research has shown that PLV use landmarks frequently in navigation \cite{szpiro2016finding}. However, visually locating and recognizing landmarks can be challenging for PLV \cite{huang2019augmented, szpiro2016finding}. For example, people with central vision loss may have difficulty seeing relatively small or low-contrast landmarks, and people with peripheral vision loss may not be able to easily locate surrounding landmarks due to limited field of view. As a result, typical landmarks for sighted people may not be visible or useful for PLV. %Additionally, sighted people define landmarks within a narrow visual cone, whereas blind people define landmarks that fall in a much broader field that include the wind, sounds and smells \cite{tsuji12005landmarks}. While PLV's visual abilities fall between those of sighted and blind people, their landmark preferences have not been adequately explored in existing research. 

Prior research has identified and categorized landmarks for both sighted people \cite{lynch1964image,yesiltepe2021landmarks,sorrows1999nature} and people who are completely blind \cite{tsuji12005landmarks,jeamwatthanachai2019indoor,wang2023understanding}. However, the visual abilities of PLV fall between sighted and blind users and may thus have different preferences in landmark selection and enhancements, revealing a knowledge gap in PLV's landmark perception. %There is thus a gap in deeply understanding how PLV select and perceive landmarks in navigation to support
As such, to enable an effective landmark augmentation system, it is important to deeply understand \textit{what landmarks PLV use} in navigation, \textit{what landmarks they prefer to augment}, and \textit{how to augment these landmarks} for PLV.%to enable an effective system that support PLV in landmark perception. 

Augmented reality (AR) technology has the potential to enhance landmarks for PLV by immersively rendering multi-modal feedback over real-world environments. Prior work for sighted users has shown that AR-based landmark systems can improve mental map development and wayfinding performance, as well as reduce mental workload compared to 2D interfaces \cite{mckendrick2016into, zhang2021enhancing}. We seek to seize this opportunity and explore how to best leverage this technology and assist PLV with landmark identification via AR augmentations. To achieve this goal, our research focuses on indoor navigation (due to the visibility limitation of AR feedback outdoors) and addresses two research questions:

\begin{itemize}
    \item  RQ 1:  How do PLV select, perceive, and utilize landmarks in navigation and mental model development?
    \item RQ 2: What are effective AR augmentations to support PLV in landmark identification?
\end{itemize}

To answer these questions, we first conduct a formative study with six PLV using contextual inquiry to understand how they perceive and use landmarks in indoor environments as well as how they prefer to enhance landmarks. %Afterwards we asked multiple questions pertaining to their mental model surrounding routes and how they chose and used landmarks, as well as exploratory and brainstorming questions about landmark-based augmentation design that they would like to have. 
The study reveals unique landmark selections for PLV. While the general landmark selection aligned with the three landmark categories for sighted people---visual, cognitive, and structural landmarks \cite{sorrows1999nature}---we identified unique subcategories for PLV, such as \textit{space with certain lighting conditions} and \textit{silhouette of an area} as visual landmarks to indicate PLV's overall impression of the space, and objects with accessibility implications (i.e., railings and elevators) as cognitive landmarks for mobility and safety purposes. Moreover, we found that PLV prefer not only augmentations on landmarks nearby but also early previews for space structure and upcoming landmarks at a distance.

Based on the formative study, we design \textit{VisiMark}, an AR interface that supports landmark perception for PLV. VisiMark includes two features: \textit{Signboards} at hallway intersections to present an overview of the hallway structure and upcoming landmarks, along with color-coded hallways (Figure \ref{fig:teaser}A), and \textit{In-situ Labels} with icons and texts to augment individual landmarks directly at their physical locations (Figure \ref{fig:teaser}B). We evaluated VisiMark with 16 low vision participants, who conducted navigation and retracing tasks indoors with and without VisiMark. %(two routes with VisiMark and two routes without). 
Participants also drew a map of the route after each navigation to demonstrate their mental model of the environment.
% A Signboard appears at intersections of hallways and presents an overview of the hallway structure and upcoming landmarks (Figure \ref{fig:teaser}A). %The hallway and landmark representations on the Signboard are in proportion to the size of the hallways and the relative positions of the landmarks. 
% It consists of arrows that represent different hallways with the arrow length and width in proportion to the hallway space. The actual hallways are also color-coded via AR to match the corresponding arrow color on the Signboard. We also label the upcoming landmarks along the hallways on the Signboard, representing the relative positions of the landmarks in the space. %icons and texts along the arrows to represent the upcoming landmarks on the hallways. The placement of icons on the signboard corresponds to the relative positions of the landmarks on the hallways. 
% Beyond the overview, we also render In-situ Labels, including icons and text descriptions, at the physical positions of the landmarks to augment the landmarks directly (Figure \ref{fig:teaser}B). Our labels are designed in vibrant colors and large sizes to ensure visibility for low vision users. The landmark icons are designed based on the landmark categories for PLV derived from the formative study. 

 %Following each navigation task, we asked them to draw a map of the path they walked and then retrace it. We gathered their feedback quantitatively and qualitatively through the navigation task and the subsequent exit questions.

Our study demonstrated the effectiveness of VisiMark. We found that VisiMark enabled PLV to perceive landmarks that they preferred but could not easily perceive before (e.g., an elevator located on the side of the user's blind eye, recessed landmarks like bathrooms). As a result, VisiMark changed participants' landmark selections from visually-obvious objects to cognitive landmarks that were more memorable and meaningful. We thus derived a new taxonomy of \textit{prefer-to-augment landmarks} in AR contexts, including unique but not visually obvious landmarks, visually challenging but cognitively important landmarks, and out-of-field landmarks that indicate potential dangers.%We also identified design implications for landmark augmentations in AR for PLV.
% We also found that PLV prefer different types of landmarks to be augmented in different stages. Visually obvious landmarks should only be augmented in preview, while common yet important facilities and object affordance are more preferred to be augmented only in situ.
% \yuhang{may need to add more interesting findings... your current findings are vague (especially those i commented off below), hard to be connected to specific measures, and also not interesting. This may be because that you don't have any significant results, but it does not mean that you want to communicate things vaguely and also does not mean that you don't have interesting results. For example, what are the landmarks PLV prefer to augment vs. they can perceive by themselves.}

% improved PLV’s landmark perception, enabling them to better recognize and memorize landmarks, and helped low vision users select important landmarks more efficiently.
 % participants completed the map drawing and retracing tasks more effectively and comfortably with VisiMark. We also found that 
%improved mental map building. 
%Low vision participants would like to augment all landmarks that meet their criteria, and visually challenging but important landmarks, especially dangers such as obstacles out of their central view. We also found their preferences for what should be augmented in preview and in situ only, along with suggestions for refining system design and offering more customization options.

In summary, we contribute the first exploration of the landmark perception of PLV as well as the design and evaluation of VisiMark, an AR system that enhances landmarks for PLV in indoor navigation. Our study reveals PLV’s needs for visual landmark augmentations and sheds light on future landmark-based navigation systems in AR for PLV.
\section{RELATED WORK}
Our work builds on prior research that defines landmarks for sighted and blind users, develops navigation assistance systems, and creates landmark enhancement AR technologies. We introduce them below to contextualize our work.
% \subsection{Current methods to facilitate cognitive map formation for people with low vision / blind}
% % not in ar
% % in ar

% \subsection{Current methods to facilitate navigation for people with low vision / blind}
% % not in ar
% % in ar
% turn by turn 
% landmark based
% navigation system in open areas (more recent research, like airport)
% Landmarks are important elements in the environments \cite{lynch1964image} 
% \subsection{Landmark-based system in augmented reality} %In summary, landmarks have a wide range of applications in navigation.
% landmarks established for sighted people are not sufficient for visually impaired people.  The authors invited 3 visually impaired participants and 3 sighted participants, but did not report more detailed differences in visual landmark choices. They also found that visually impaired people tend to describe landmarks from a subjective egocentric point of view, in contrast to the objective static physical fixtures. 

\subsection{Landmark Definition, Categories, and Usage}\label{Landmark Definition, Categories, and Usage}
Prior research has explored landmark definitions and usage for sighted people. Landmarks are defined as stationary, distinct, and salient objects or places in the space that are more likely to be selected as spatial references \cite{lynch1964image,millonig2007developing}. Landmarks serve multiple purposes throughout different stages of wayfinding \cite{yesiltepe2021landmarks}. Previous work found that people use landmarks to orient themselves \cite{downs2011cognitive,philbeck2005remembered}, locate destinations \cite{klippel2005structural}, guide others \cite{raubal2002enriching,tom2003referring, lovelace1999elements}, and construct mental representations of unfamiliar environments \cite{michon2001and}.

Prior work summarized the main criteria for landmarks for sighted people as permanence, uniqueness, and identifiability \cite{yesiltepe2021landmarks}. From the aspect of landmark saliency, Sorrows and Hirtle \cite{sorrows1999nature} classified landmarks into three categories: visual, structural, and cognitive landmarks. Visual landmarks are objects with salient visual features (i.e., shape, color,  size), cognitive landmarks are those with important meanings (i.e., culturally, historically, or personally important), and structural landmarks are the ones that take a vital role or location in the structure of the space. From the aspect of visibility, landmarks can be categorized into two types: global and local landmarks \cite{steck2000role,lynch1964image,yesiltepe2021landmarks}. 
Global landmarks are visible from many angles and distances \cite{lynch1964image}, while local landmarks can only be seen from up close \cite{steck2000role}.

Some research has investigated how blind users define and categorize landmarks. Different from sighted people, blind people perceived landmarks across multiple sensory channels, including tactile, auditory, smell, and sense of mass \cite{jeamwatthanachai2019indoor, wang2023understanding, tsuji12005landmarks}. For example, Tsuji et al. \cite{tsuji12005landmarks} found that sounds and smells were noticed more by blind people than by sighted people. 
Saha et al. \cite{saha2019closing} divided landmarks into five categories for blind people: structural, sound, tactile, air, and smell, with structural landmarks being shared by both blind and sighted people. Cognitive landmarks are barely used by blind people since it is difficult for them to detect and recognize semantic features \cite{sato2019navcog3}. Wang et al. \cite{wang2023understanding} further summarized three criteria of landmarks for blind people: permanent, reliable, and identifiable, highlighting that landmarks used by blind people need to be fixed and easy to find via tactile feedback. % and they frequently used tactile feedback to detect landmarks. 

Despite the thorough research for sighted and blind people, little attention has been drawn to PLV. Recently, a couple of recent projects have started involving some low vision participants and identified preliminary findings \cite{wang2023understanding,tsuji12005landmarks}. For example, Tsuji et al. \cite{tsuji12005landmarks} found that PLV often paid attention to visual objects that sighted people rarely notice, such as the texture underfoot. However, they investigated blind and low vision users altogether, overlooking the unique needs of PLV, whose visual abilities fall between sighted and blind people. No research has deeply explored the landmark selection for PLV and their preferences for landmark augmentations. Our research will thus fill this gap to thoroughly understand PLV's landmark perceptions.

% However, the authors only provided examples and did not categorize the differences in visual landmark choices between low vision and sighted users. 

% Although some existing works \cite{wang2023understanding,tsuji12005landmarks} involved low vision participants in their interviews about how low vision and blind people identify landmarks, they did not report PLV and blind people separately. Unlike blind people, PLV rely on their remaining functional vision heavily everyday. There is a research gap in how PLV perceive and leverage landmarks in their wayfinding and mental model construction.

% How blind people utilize landmarks are similar to sighted people. 

\subsection{Navigation Technology for Blind and Low Vision People}

Many navigation systems have been designed to support blind and low vision people in wayfinding. Most of them focused on turn-by-turn instructions for blind users via auditory \cite{gaunet2006verbal, helal2001drishti, ahmetovic2016navcog, simoes2016blind} and haptic feedback \cite{ertan1998wearable, azenkot2011smartphone, flores2015vibrotactile, liu2021tactile}. For example, NavCog \cite{ahmetovic2016navcog} provided audio turn-by-turn instructions via smartphones and utilized real-time localization to inform blind users about nearby points of interest and accessibility issues. Liu et al. \cite{liu2021tactile} developed Tactile Compass, a hand-held device that generated directional tactile feedback with a rotatable needle pointing towards the planned direction. Some haptic feedback systems also incorporated robotic assistants \cite{lacey1998application, kulyukin2005robocart, guerreiro2019cabot} to guide blind users in navigation.

Some turn-by-turn instruction systems are designed specifically for PLV by providing visual guidance \cite{chi2022enabling, zhao2020effectiveness,yang2021lightguide}. For example, Stern et al. \cite{stent2010iwalk} presented iWalk, a speech-enabled local search and navigation smartphone application for PLV, providing real-time turn-by-turn walking directions in speech and high-contrast text. Zhao et al. \cite{zhao2020effectiveness} explored the design of wayfinding guidance in AR for PLV, comparing the visual and audio feedback. They found that PLV could better perceive length information with lower cognitive load and make fewer mistakes using visual feedback compared to audio feedback. Moreover, %Chi et al.  built an AR application that provided both visual and audio instructions for PVL but have not conducted user studies.
Yang et al \cite{yang2021lightguide} developed LightGuide, which indicated the direction of travel via the position of a light within users’ visual field for people with light perceptions. 

In addition to turn-by-turn instructions, several systems have been developed to support safe navigation by detecting obstacles for PLV \cite{bai2017smart, zhao2019designing, da2021wearable, fox2023using}. For example, Bai et al. \cite{bai2017smart} detected obstacles along the way and explored the feasibility of different feedback to inform walkable directions, such as visual indicators (e.g., position of a virtual circle) and speech alerts of obstacles. Zhao et al. \cite{zhao2019designing} supported stair navigation for PLV by highlighting stair edges and railings via projection-based and smartglass-based AR. Lately, Fox et al. \cite{fox2023using} also explored the effectiveness of different AR cues to augment obstacles for PLV, finding that 3D world-locked AR cues were superior to directional heads-up cues.

Beyond wayfinding and safe navigation, it is also important for PLV to explore the environment actively \cite{chrastil2015active} and build mental models with landmark knowledge \cite{siegel1975development,kim2021acquisition}. However, to our knowledge, no research has investigated how to enable PLV to better locate and recognize landmarks in navigation.

\subsection{Landmark Enhancements in Augmented Reality}

Researchers have developed landmark enhancement systems in AR for sighted people. Zhang et al. \cite{zhang2021enhancing} found that the AR navigation system with landmarks enhanced users' mental map development and improved wayfinding performance. Liu et al. \cite{liu2021spatial} used AR icons to label semantic landmarks, showing that virtual semantic landmarks can promote the acquisition of spatial knowledge. Additionally, Zhu et al. \cite{zhu2022personalized} developed a system that adaptively selected and augmented landmarks based on landmark saliency and users' familiarity with the environments.
Uniquely, Qiu et al. \cite{qiu2023navmarkar} investigated how to design landmark-based navigation systems tailored for older adults, including a 3D map of the floor plan, turn-by-turn instructions that indicated the next landmark, and an interactive board attached to each landmark to enable users to acquire more information.%They designed interactive landmark information boards at their physical locations,  Their designs focused on integrating additional landmark information with turn-by-turn instructions, but might result in too much text to read when applied to PLV.  \yuhang{no...i mean suggesting what? if they design for older adults, did you use any new designs?}. 

For blind people, existing landmark enhancement systems have focused on providing auditory feedback \cite{balata2016automatically, balata2018landmark, may2020spotlights, fiannaca2014headlock, coroama2003chatty}. For example, Balata et al. \cite{balata2016automatically, balata2018landmark} designed a system that generates landmark-enhanced navigation instructions via audio feedback, revealing that users preferred the landmark-enhanced instructions than conventional turn-by-turn instructions. May et al. \cite{may2020spotlights} developed an auditory environment in mixed reality to simulate a virtual space, presenting landmarks through spatial audio to facilitate mental map formation.

For PLV, while prior research has demonstrated the potential of AR head-mounted displays as a powerful accessibility support \cite{zhao2017understanding, min2021augmented}, no AR augmentations have been designed to support them in landmark perception. The most relevant research is by Huang et al. \cite{huang2019augmented}, who designed an AR system that recognized and augmented text with high-contrast letters and text-to-speech to assist PLV with sign reading. However, they only focused on sign augmentation without considering other types of landmarks. In contrast to prior research, we deeply understand and categorize landmarks for PLV and design VisiMark that augments various landmarks via both AR previews and in-situ labels.

\section{FORMATIVE STUDY}
To fundamentally understand what landmarks PLV commonly use and how they perceive and leverage landmarks in navigation (RQ 1), we first conducted a formative study with six low vision participants using the method of contextual inquiry \cite{karen2017contextual}. The study identified the unique landmarks for PLV and provided insights on how to augment landmarks for them, thus inspiring the design of VisiMark.

%Our research involved minimal risks and received IRB approval to conduct human subjects studies.

\subsection{Methods}
% Following a user-centered design approach \cite{abras2004user}, we conducted a contextual inquiry study \cite{karen2017contextual} with 6 low vision participants to investigate how PLV perceive and use landmarks during their navigation in indoor environments. 

\subsubsection{Participants}
We recruited six participants (P1-P6, 2 females and 4 males), whose ages ranged from 33 to 78 ($Mean = 62.7$, $SD = 15.5$). Two participants (P1, P2) had prior AR experiences. Three participants (P1, P3, P6) were legally blind (i.e., best-corrected visual acuity in the better eye worse than 20/200 or visual field narrower than 20 degrees) but had functional vision to navigate. All participants wore eye glasses. Table \ref{tab:demographics_formative} shows participants’ demographic information. We recruited participants via email lists and non-profit organizations. A participant was eligible if they were at least 18 years old and had low vision but were able to use functional vision in daily activities. Participants were screened via phone or email to ensure they met these criteria. Participants were compensated \$20 per hour and were reimbursed for travel expenses. \colorchange{This study was approved by the Institutional Review Board (IRB) at our university.}

%TC:ignore
\begin{table*}[t]
\scriptsize
\centering
\begin{tabular}{>{\centering\arraybackslash}p{0.2cm}>{\centering\arraybackslash}p{1.3cm}>{\centering\arraybackslash}p{1.9cm}>{\centering\arraybackslash}p{1.3cm}>{\centering\arraybackslash}p{2cm}>{\centering\arraybackslash}p{1.9cm}>{\centering\arraybackslash}p{4.1cm}>{\centering\arraybackslash}p{2.2cm}}
%\Xhline{2\arrayrulewidth}
\toprule
% \multirow{2}{*}{\textbf{ID}} & 
% \textbf{Age/} & 
% \multirow{2}{*}{\textbf{Diagnosis}} & 
% \textbf{Legally} & 
% \multirow{2}{*}{\textbf{Visual Acuity}} & 
% \multirow{2}{*}{\textbf{Field of View}} & 
% \multirow{2}{*}{\textbf{Other Visual Difficulties}} & 
% \textbf{Prior AR} \\
% & \textbf{Gender}& &\textbf{Blind} & & & & \textbf{Experiences}\\
  \textbf{ID} & \textbf{Age/Gender} &  \textbf{Diagnosis} & \textbf{Legally Blind} &  \textbf{Visual Acuity} &  \textbf{Field of View} &  \textbf{Other Visual Difficulties} & \textbf{Prior AR Experiences} \\
% \Xhline{2\arrayrulewidth}
\hline
P1 & 67/M & Not know & Y & 20/2400 \& 20/2800 & Full & Sensitive to light &  Y\\
\hline
P2 & 78/M & Cone dystrophy & N & Not know & Full & Sensitive to light; cannot tell color shades& Y \\
\hline
\multirow{2}{*}{P3} & \multirow{2}{*}{60/F} & \multirow{2}{*}{Spinal meningitis} & \multirow{2}{*}{Y} & \multirow{2}{*}{R: 20/400; L: 20/2200} & \multirow{2}{*}{Full} & Sensitive to light; depth perception issue; & \multirow{2}{*}{N} \\
& & & & & & confused between black and blue& \\
% P3 & 60/F & Spinal meningitis & Y & R: 20/400; L: 20/2200 & Full & Sensitive to light; confused between black and blue; depth perception issue & N \\
\hline
P4 & 68/F & Macular degeneration & N & Not know & Central vision loss &  N/A & N \\
\hline
P5 & 70/M & Macular degeneration & N & R: 20/50; L: 20/50 & Full & N/A&  N\\
\hline
P6 & 33/M & Retinitis pigmentosa & Y & R: < 20/200; L: < 20/200 & Peripheral vision loss & Sensitive to light; depth perception issue & N \\
\bottomrule
\end{tabular}
\caption{Participant demographics in the formative study.}
\label{tab:demographics_formative}
  % \vspace{-5ex}
\end{table*}
%TC:endignore

\subsubsection{Procedure}
We conducted a single-session contextual inquiry study in an indoor environment. The study lasted approximately two hours. We started the study with an initial interview, asking about participants' demographics, visual conditions, and familiarity with AR. We also asked about participants' prior experiences with navigation and landmark recognition, such as the technologies they use for navigation, whether they use or try to memorize any landmarks in navigation, and what types of landmarks they commonly use.% (e.g., Do you use any technology to navigate regularly? Do you use landmarks during navigation? If yes, what types of landmarks?). 

Participants then conducted navigation tasks along two indoor routes. Both routes featured various potential landmarks, such as doors, stairs, water fountains, emergency resources, and hallway decorations. Meanwhile, the two routes presented different characteristics: one route was an open area (i.e., lobby) with more environmental dynamics (e.g., people moving around, temporary changes of tables and chairs), while the other was a hallway environment that maintained a relatively consistent setting.

For each route, participants first navigated to a destination by following verbal instructions. A researcher on the team followed the participant and provided turn-by-turn instructions when the participant approached a decision point (i.e., an intersection point where participants decided whether to turn and the turning directions). During the navigation, participants thought aloud, talking about what landmarks they saw and why they would use them as landmarks. The destination and starting point were designed to be the same location for both routes. After returning to the start, participants were asked to retrace the route to the destination again independently without any instructions. During the retracing, participants also thought aloud to discuss how they memorized the route and how they determined when and where to turn along the route. During the navigation and retracing, we observed participants' behaviors (e.g., what directions or objects they looked at or pointed to), took notes of their thinking-aloud contents, and asked follow-up questions, such as the challenges participants faced when perceiving certain landmarks, their strategies in choosing landmarks, and the rationales.

% where we observed their navigating through two different indoor routes. After the initial navigation of each route with direction instructions, we asked participants to describe the route they had just navigated and then retrace the route independently without any instructions. During the navigation, we observed participants' behaviors  and asked what objects or elements they used as landmarks, the challenges of seeing landmarks, their strategies in choosing landmarks, and the rationales. 

We ended the study with an exit interview, asking participants to summarize different types of landmarks they used as well as the importance of landmarks in wayfinding and mental map development tasks (Likert scale from 1 to 5, where 5 means most important). Lastly, participants brainstormed ideas for suitable methods to augment landmarks. \colorchange{Detailed interview questions can be found in Appendix \ref{Interview Questions for the Formative Study}.} %augmentations. Lastly, we asked participants a few exit interview questions, including rating the importance of landmarks in wayfinding and building a mental map from 1 to 5, where 1 meant least important and 5 meant most important. 

\subsection{Analysis}
We video-recorded the study and transcribed the audio data using an automatic transcription tool. We also manually went over the transcripts to correct any transcription errors and inserted our observations (e.g., participant behaviors) into the corresponding positions in the transcripts. We analyzed the transcripts using thematic analysis \cite{braun2006using,clarke2017thematic}. Three researchers open-coded four participants' data independently and collaboratively developed a codebook through discussions to resolve any discrepancies. We then developed themes and sub-themes from the codes by clustering relevant codes using axial coding and affinity diagrams. Once initial themes and sub-themes were identified, researchers cross-referenced the original data, the codebook, and the themes to make final adjustments, ensuring all codes were correctly categorized.

\subsection{Findings}
% In this section, we report how low vision participants defined, categorized, and used landmarks in their navigation and mental map building. We also investigate the challenges they face and brainstorm potential landmark augmentations.

\colorchange{Our study revealed that landmarks played an important role in both wayfinding and mental map development for PLV, with mental map development receiving a slightly higher importance rating ($M=4.67$, $SD=0.52$) than wayfinding ($M=4.17$, $SD=0.52$) but no significant difference ($t_5 = 1.94$, $p = .111$)}. Similar to sighted people \cite{downs2011cognitive,klippel2005structural, tom2003referring, lovelace1999elements}, we found that PLV also identified landmarks visually to find specific locations (P3, P6), recognize the environment and locate themselves (P3, P5, P6), and give instructions to others (P4). Most participants (P1, P3, P4, P6) stated that landmarks were more important in unfamiliar places and they would look for landmarks when visiting a place for the first time. However, different from sighted people, we identified unique types of landmarks adopted by PLV as well as the challenges they faced when locating and perceiving landmarks, which inspired the design of our landmark augmentation interface. We elaborate on our findings below.

% \paragraph{Familiarity}
% Most participants (P1, P3, P4, P6) stated that they would look for landmarks when visiting a place for the first time, while P5 was more inclined to ask others for directions. Whether familiarity changes landmark selections varies among individuals. P1 and P3 believed their landmarks would change with increased familiarity. As they became more familiar with the place, P1, P3, and P6 mentioned that they would rely on dead reckoning instead of using landmarks. In contrast, P4 and P5 claimed that familiarity does not affect their landmark selection, as they incorporate landmarks into their mental maps. These comments suggest that landmarks are more important in unfamiliar places.

\subsubsection{Landmark Characteristics for Low Vision}\label{Landmark Characteristics for Low Vision}
While echoing prior research for sighted and blind users that landmarks should be \textit{unique} and \textit{permanent} \cite{yesiltepe2021landmarks,wang2023understanding}, we found that PLV employed different features to determine landmarks due to their visual abilities. We summarize the landmark characteristics for PLV below.

\textit{\textbf{Prefer landmarks in visual modality.}}\label{Prefer landmarks in visual modality.} Despite experiencing vision loss, low vision participants relied primarily on visual cues to identify landmarks in navigation. Most landmarks they used were visual, including visually obvious and structurally salient features (categorizations in Section \ref{Landmark Taxonomy for Low Vision.}). Meanwhile, some participants also paid attention to landmarks in other modalities. Two participants (P3, P5) noticed smells (e.g., chemical smells from a lab) and sounds (e.g., music) on their routes. %confirming prior finding that PLV tended to notice sounds and smells more than sighted individuals \cite{tsuji12005landmarks}. 
However, all participants preferred identifying landmarks via visual cues, with P6 noting a preference for ``focusing more on things [he] can see than feel or [hear].'' P2 explained that he found it challenging to accurately locate the landmark with sound alone. % ``[As for the] hear stuff, I wouldn't be able to identify their relative [position].''%\yuhang{need more about why PLV prefer visual more than other modalities... i remembered that you had them before. Otherwise, we are not highlight the theme of this section.} 

%\textit{\textbf{Leverage unique object combinations as landmarks.}} Instead of individual items, many low vision participants (P2-P6) combined multiple objects to form landmarks. Since PLV often focused on visually obvious objects, selecting as many unique landmarks as sighted individuals can be challenging. As a result, low vision participants tended to use unique combinations of objects as landmarks. For example, P2 combined a water fountain and a nearby eye wash station as a landmark, and P6 used the combination of a hand sanitizer with a statue as a landmark. P3 further pointed out that selecting unique combinations of landmarks would help prevent confusion caused by repeated elements. 
%\yuhang{this sounds to be one object... a ramp with a railing? then it's not a combination}%Although P3 commented that "[I would not choose] ones (landmarks) that are really close together. They have to have some distance.", she agreed that she would pick a unique combination of landmarks to avoid confusion with things appearing multiple times. Since low vision participants faced difficulty distinguishing similar objects, a unique combination of objects could be easier for them to identify.
%\yuhang{need a stronger evidence then what i commented off.}
%In a relatively similar environment, low vision participants chose to use a unique combination of landmarks to satisfy the prerequisite of landmark uniqueness. This approach likely stems from their visual abilities, as they may struggle to identify as many unique landmarks as sighted individuals in similar settings.

\textit{\textbf{Rely on consistent landmark appearance.}}
While sighted and blind people commonly interpret permanent landmarks as not movable \cite{burnett1998turn,wang2023understanding}, our low vision participants (P3, P4, P5) stressed that the appearance of landmarks should also be consistent over time. We found that varying states of the same object could cause concerns for PLV, being perceived as different landmarks. As P3 commented on a shirt store, ``The shirts might be put in a shop or something [when the shop is open]. [But] when [the shop] closes, it looks different.'' Due to their visual conditions, PLV held a particularly high standard for object permanence when selecting landmarks.% While permanence for both sighted and blind users generally refers to landmarks being consistently located , PLV also require that the appearance of the landmarks remains consistent.

\subsubsection{Landmark Taxonomy for Low Vision.}\label{Landmark Taxonomy for Low Vision.}
According to Sorrows and Hirtle \cite{sorrows1999nature}, landmarks can be classified into three categories for sighted people---visual, cognitive, and structural landmarks. We found that landmarks for PLV aligned with these categories, but they included more unique subcategories due to PLV's visual abilities and needs. Note that these categories are not mutually exclusive; many landmarks can fall into multiple categories.%\yuhang{if possibel, it would be nice to add a row of photos to show the unique landmarks for PLV. Refer to my stair paper at ASSETS 2018.}

%TC:ignore
\begin{figure*}[t]
    \centering
    \begin{subfigure}{0.99\textwidth}
        \centering
        \includegraphics[width=\textwidth]{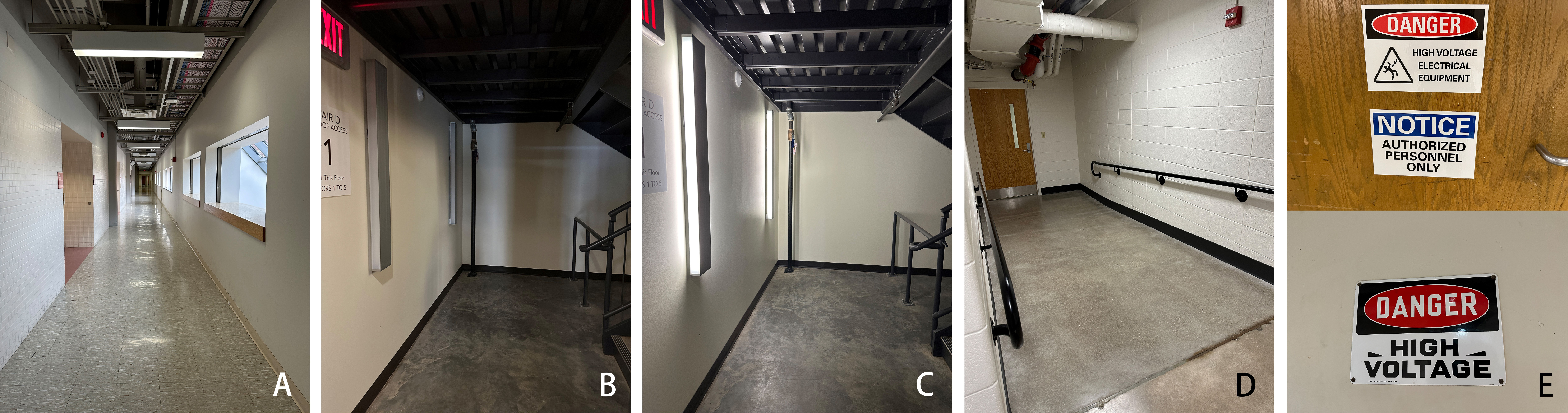}
    \end{subfigure}
    \caption{(A) A bright hallway with a reflective floor, where shadows on the floor look like a ladder; (B-C) Staircases under different lighting conditions: dark vs. bright; (D) A ramp with railings as a landmark; (E) Danger signs as landmarks.} 
    \Description{This image shows unique landmarks for PLV. Image (A) A bright hallway with a reflective floor, where shadows on the floor look like a ladder; (B-C) Staircases under different lighting conditions: dark vs. bright; (D) A ramp with railings as a landmark; (E) Danger signs as landmarks.}
    \label{fig:landmark_examples}
\end{figure*}
%TC:endignore

%TC:ignore
\begin{figure*}[h]
    \centering
    \begin{subfigure}{0.99\textwidth}
        \centering
        \includegraphics[width=\textwidth]{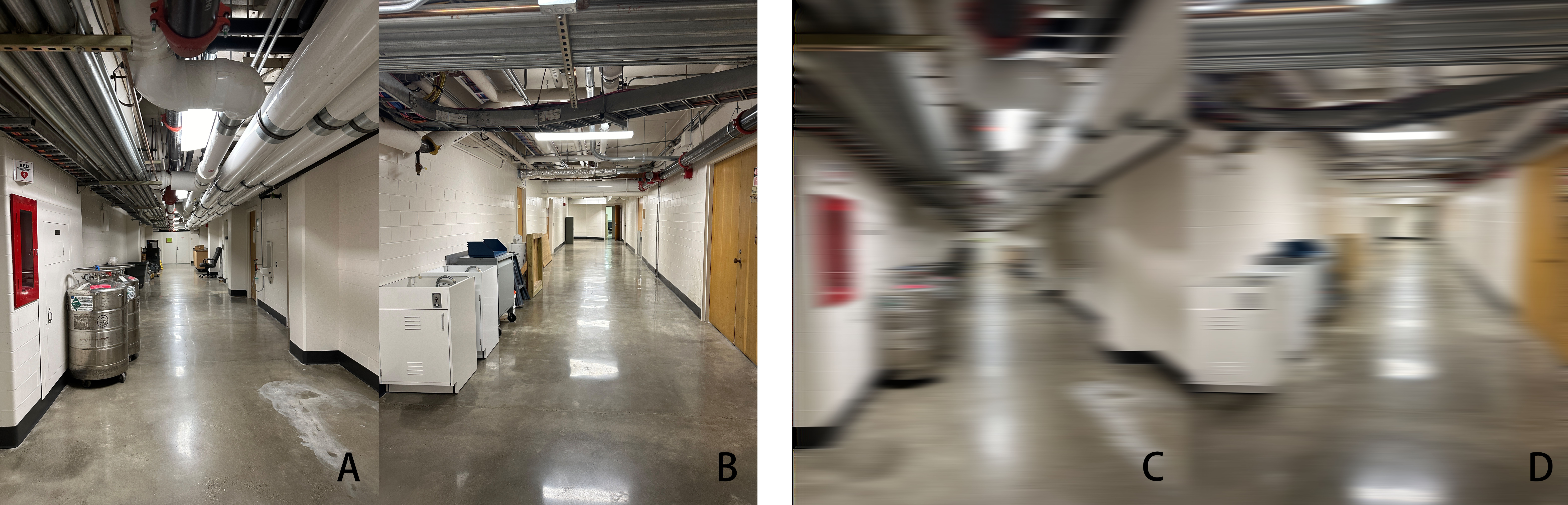}
    \end{subfigure}
    \caption{Perceived ``silhouette'' of an area: (A-B) Two similar hallways. (C-D) Blurred to simulate how the ``silhouette'' of an area looks to PLV, where similar hallways have different color blocks.} 
    \Description{This image shows perceived ``silhouette'' of an area: (A-B) Two similar hallways. (C-D) Blurred to simulate how the ``silhouette'' of an area looks to PLV, where similar hallways have different color blocks.}
    \label{fig:silhouette}
\end{figure*}
%TC:endignore

\textbf{\textit{Visual Landmarks.}}
Similar to sighted people \cite{sorrows1999nature}, PLV chose landmarks based on visual characteristics, which included large size (P1-P6), vibrant colors (P1-P6), high contrast (P1-P6), special textures (e.g., tiles, P2-P5), and protruding features (e.g., clocks, P2-P6). Beyond individual objects, we found that PLV were sensitive to the \textbf{\textit{lighting conditions}} in the environment (Figure \ref{fig:landmark_examples}A-C). All participants noticed the changes in lighting conditions and used them as landmarks. For example, they picked dark stairs (P1, P5, P6), dim hallways (P2), and reflective floors (P3, P4) as landmarks. These extreme lighting conditions (e.g., too bright or too dark) can significantly influence participants' visual perception and cause safety concerns, thus leaving them a deep impression. P3 explained why she noticed the reflective floor, ``All those lights are over [the floor]. It makes me think that there's something like a ladder [because of the shadows].'' 

Moreover, we also observed that beyond individual objects, PLV used the \textbf{\textit{perceived ``silhouette'' of an area}} as landmarks without discerning the specific objects (P1, P4, P5), as shown in Figure \ref{fig:silhouette}. For example, P1 mentioned that he remembered more about the hallway as a whole rather than specific objects within it, as he could not identify unique objects in the hallways. Similarly, P5 couldn't identify the specific items in the hallway, but he could differentiate between hallways based on the quantity of objects along the hallways.

In summary, the range of visual landmarks for PLV expands from individual objects to broader areas, including both lighting conditions and the overall structure or silhouette of a space.

\textbf{\textit{Cognitive Landmarks.}}
Aligned with previous research for sighted people \cite{yesiltepe2021landmarks}, PLV also selected landmarks considering their cognitive or semantic meanings. Most participants (P1-P5) selected landmarks because of personal experiences. For example, P2 was sensitive to numbers because of requirements in his previous career. Incidents that happened previously during navigation also stood out in memory (P1, P4, P5). As P1 said, ``[If I was asked to navigate again, I would remember] I've made a wrong turn [here, so] I need to go this way.'' Similarly, P5 remembered the water fountain as he used it during his wayfinding. 

Although small prints on the signs or maps could be hard for them to read, all participants paid attention to them. P3 even stopped and read the notice on the door carefully during navigation. In particular, all participants were attentive to signs indicating \textbf{\textit{danger and emergency}}, such as danger signs on the door (see Figure \ref{fig:landmark_examples} E) and emergency exits.

Elements with important functions are another typical types of cognitive landmarks used by our participants, such as restaurant (P2, P3, P5, P6), front desk (P4, P5, P6), and bathrooms (P2, P3, P5, P6). Remarkably, low vision participants paid additional attention to \textbf{\textit{landmarks with accessibility purposes}} (Figure \ref{fig:landmark_examples} D), including railings (P2-P6), ramps (P1-P6), and elevators (P2-P6), so that they can use them as mobility aids. Participants were highly sensitive to surface level changes (e.g., stairs, ramps) that can cause potential danger \cite{zhao2018looks}, resulting in memorizing them as landmarks. As P5 explained, ``I think I'm still focusing more on a [ground] level because the ground, as you know, that's a safety thing. [To avoid] falling, you have to really make sure that your ground is stable. People rely on that.'' %When it becomes difficult, let's say walking in places like shaded areas, I can't rely on my visual field to see the steps... And then you if you're much more careful, take smaller steps. '' 
For the same reason, PLV paid attention to obstacles that may obstruct their path or cause them to fall and used them as landmarks.

In summary, PLV prioritized cognitive landmarks that directly relate to their safety and accessibility needs.  %This increased awareness of features that can help or hinder their mobility highlights how PLV uniquely engage with and rely on cognitive landmarks.

\textbf{\textit{Structural Landmarks.}}\label{Structural Landmarks}
 Low vision participants also used structural landmarks, such as stairs (P1-P6), pillars (P2, P3, P5), doors (P1-P6), and walls (P1, P3) during navigation. These structural elements are important because they often signal changes in the architectural layout. For example, P1 remembered a specific door because it marked the path he needed to take, while P2 recalled the open double doors that led to another area. %Some participants picked landmarks based on their \textbf{\textit{locations}}. For instance, P2 noticed equipment an electrical cord hanging down in an unusual spot, making it stand out as a landmark.
 %\yuhang{why they are structural landmarks? sound like obstacles---cognitive.. importance comes from the location is part of the structural landmark definition, I want to show PLV's criteria align with this definition here}

Moreover, our participants showed a tendency to understand \textbf{\textit{the overall structure of the floor plan}}. They frequently leveraged the \textbf{\textit{the size of the space}} as landmarks, including both length (P1, P3, P4, P5, P6) and width (P4, P5) of a hallway. For example, P4 described the hallway as ``a much more open space'' and remembered it as a long hallway. Participants (P1, P3, P4) paid particular attention to intersections because they connected different possible routes. As P3 noted, ``When I come up to an intersection of a hallway like this, I always look and see what I can see.'' This highlights the importance of presenting an overview at intersections to support navigation decisions.

In summary, low vision participants use both structural and spatial characteristics as landmarks, such as the size and overall layout of a space. Their navigation strategies emphasize understanding connections between structural elements, although their visual conditions may affect the accurate perception of these features.

\subsubsection{Challenges and Desired Augmentations for Landmarks.} 

Low vision participants identified several challenges and desired enhancements for landmarks. We elaborate on their needs and preferences as design guidelines (DG 1-4) below to guide the design of VisiMark.

\textbf{\textit{Enhancing landmarks visually (DG 1).}} Since low vision participants predominantly prefer landmarks in the visual modality (see Section \ref{Landmark Characteristics for Low Vision}), visual augmentations should be prioritized. Although many participants (P1-P5) acknowledged the potential benefits of incorporating audio augmentation, some expressed reservations. P3 emphasized that audio feedback would not be as timely as visual cues, especially for those who walk quickly. Meanwhile, P4 was worried about potential disruption to others, noting that audio feedback could result in ``too much noise around.'' Given these concerns, visual augmentations appear to be more suitable for enhancing landmarks.

\textbf{\textit{Highlighting hallway structures (DG 2).}} Low vision participants were attentive to the overall layout and size of spaces but faced difficulty in accurately perceiving such information. For example, P3 made turning decisions based on whether a hallway seemed to lead to a dead end but misrecognized a hallway as a dead end. %saying, ``I would say don't go that way, because that looks like the end of the hallway,'' even when it wasn’t a dead end. 
She also found it difficult to distinguish between flat spaces and ramps due to the lack of depth perception. %which further complicated navigation. 
%It is thus important to augment hallway structures for PLV, including dead ends.

Participants offered suggestions for improving hallway navigation. P5 proposed painting the hallways in different colors, comparing it to how it’s done in schools: ``You [know that you] are in the red hallway, as opposed to trying to tell a five-year-old [to] go down to the third hallway [and] turn left.'' Similarly, P1 suggested highlighting the edges of the hallways, ``[highlighting the edges of the hallways] would help. [It] might also help to provide directions related to distance, so you could watch as it changes when you get closer or farther. That would help.'' Overall, it is important to augment hallway structures and space characteristics for PLV.
%participants expressed a preference for having more clues within hallways to aid in memorization and depth perception.

\textbf{\textit{Converting local landmarks to global landmarks (DG 3).}}\label{Converting local landmarks to global landmarks.} As discussed in Section \ref{Structural Landmarks}, some participants would stop at the intersections to look for landmarks ahead. However, due to visual acuity challenges, participants (P2, P3, P4, P5) reported difficulty seeing the landmarks clearly from a distance, which limited their ability to use them as global references. To address this, several participants expressed a need to be aware of upcoming landmarks, especially when their view was ``blocked off'' (P3, P6) or ``crowded'' (P3, P5, P6). They also suggested that making landmarks noticeable from different angles would be beneficial, consistent with the definition of global landmarks (see Section \ref{Landmark Definition, Categories, and Usage}). As P6 remarked, ``projecting [a flat landmark] to be 3D or something would probably be helpful.'' As a result, providing an overview with upcoming landmarks could be a useful augmentation for PLV.
%Additionally, participants tended to to focus at eye level and on the floor (P2-P6), recommending lowering elements positioned too high (P3, P4, P5) and indicating objects below eye level (P3, P6) to enhance visibility and usability. 
%\yuhang{need adjustment... see overleaf comments}

\textbf{\textit{Augmenting landmarks in-situ (DG 4).}} All participants expressed a desire to enlarge existing texts and landmarks in the environment, with most (P1-P4) advocating for large text labels on small objects (P1) or room numbers (P4). Participants also suggested coating or outlining landmarks with a bright color. As P5 mentioned, ``coating [landmarks] out with a bright red or something, that would be certainly [helpful], you know, that would be a landmark all the time for me.'' 
%All participants agreed that highlighting the landmarks with contours was a good idea. As P6 explained, ``In terms of outlining [landmarks], if there's a specific color to that, you think [the landmark] stands out the most''. 
However, P4 preferred adding icons, noting that ``the outline for me is not better than an icon.''

\section{VISIMARK: LANDMARK AUGMENTATION INTERFACE DESIGN FOR LOW VISION}
%\yuhang{need to add the goal of this design. We should say something that we design an AR interface with augmentations to (1) support low vision people xxx and (2) use the interface as design probes to further understand how to best design landmark augmentations.---so that we set up a foundation for later argument that we are still highlighting all kinds of landmarks and then evaluate what to augment} 
Based on findings from the formative study, we designed \textit{VisiMark}, a landmark augmentation interface in AR to support PLV in landmark perception. Moreover, following the user-centered design approach, we will use VisiMark as a design probe to investigate \textit{what to augment} and \textit{how to augment} landmarks for PLV to inspire future navigation systems (RQ 2). We elaborate on the landmark augmentation design \colorchange{and system implementation} below.

\subsection{VisiMark: AR Augmentations for Landmarks}

\colorchange{We designed VisiMark to be a head-mounted AR system. We chose head-mounted AR as it provides hands-free interactions, which can be particularly beneficial to low vision users who hold other assistive support (e.g., white cane, guide dog) in navigation \cite{thomas2009wearable}. Compared to other AR platforms (e.g., mobile AR), head-mounted AR also mitigates attention switching between the real world and an additional display, enabling users to focus on the surrounding environments and facilitating safety \cite{zhao2019designing}.} Following the guidelines, VisiMark provides visual augmentations to enhance landmarks (\textit{DG 1}). With the needs of both overview (\textit{DG 2\&3}) and in-situ augmentations (\textit{DG 4}), our interface offers two design components: \textit{Signboards} at intersections for preview and \textit{In-situ Labels} for visibility of individual landmarks. %We describe the two design components below.% first describe how to select landmarks in VisiMark, followed by a detailed introduction to the design.

\subsubsection{Signboards: Overview of Hallway Structures and Upcoming Landmarks}

\begin{figure*}[t]
    \centering
    \begin{subfigure}{0.99\textwidth}
        \centering
        \includegraphics[width=\textwidth]{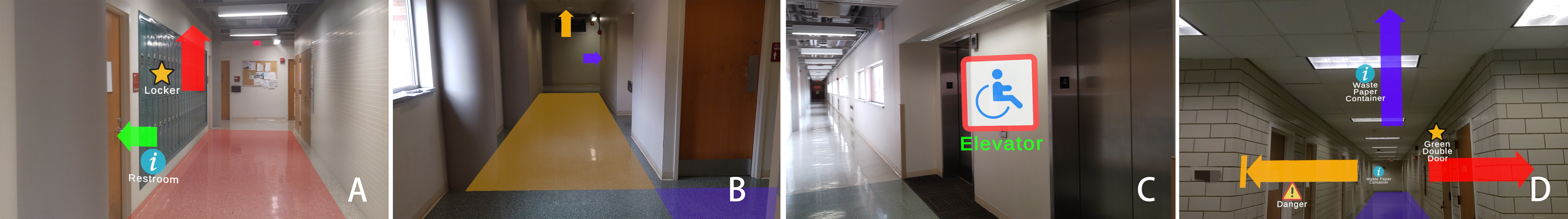}
    \end{subfigure}
    \caption{VisiMark design features: (A) Signboards that provide an overview of hallway structures and upcoming landmarks; (B) Color-coded hallways and their virtual representations (i.e., arrows) on the Signboard; (C) In-situ Labels with icons and texts for individual landmarks. (D) The dead-end marker at the tip of an arrow to represent the left hallway is a dead end.}
    
    \Description{This image shows VisiMark design features: (A) Signboards that provide an overview of hallway structures and upcoming landmarks; (B) Color-coded hallways and their virtual representations (i.e., arrows) on the Signboard; (C) In-situ Labels with icons and texts for individual landmarks. (D) The dead-end marker at the tip of an arrow to represent the left hallway is a dead end.}
    \label{fig:VisiMark}
\end{figure*}

We designed \textbf{\textit{Signboards}} to provide a preview of overall hallway structures and landmarks (\textit{DG} 2 \& 3) to support PLV's turning decisions at intersections (Figure \ref{fig:VisiMark}A). \colorchange{Signboards are anchored at each hallway intersection and remain stable as users change their viewing orientation and position.} They illustrate the hallway structures by using virtual arrows to represent each hallway. The arrow dimensions (i.e., length and width) were proportionate to the hallway dimensions. Moreover, we put a dead-end marker (i.e., a vertical line) on the tip of an arrow to indicate a hallway being a dead end (\textit{DG 2}, Figure \ref{fig:VisiMark}D). The physical hallways are color-coded with virtual overlays, matching the color of their arrow representations on the Signboard (Figure \ref{fig:VisiMark}B). The color-coded design allows PLV to use the hallways themselves as landmarks, and the virtual overlay also enables them to better perceive the hallway edges \colorchange{and endpoints}. %This design aimed to provide PLV, particularly those with depth perception issues, with a clear understanding of upcoming hallway structures. 
% Additionally, we implemented color-coded hallways to provide more cues for memorization beyond simple directional terms like "left" and "right," and to help users better perceive the edges of the hallways.

On the Signboard, we also labeled the upcoming landmarks along each hallway with icons and texts (\textit{DG} 3). The positions of the icons on the arrows indicated the relative positions of the landmarks in the physical hallway. %\colorchange{For example, in Figure \ref{fig:VisiMark}D, the green double door icon is positioned closer to the intersection than the danger sign, indicating their respective physical distances.} 
This feature was designed to provide a preview of distant landmarks before PLV entered the hallways. To suitably present the landmarks, we designed landmark icons based on the landmark categories derived from the formative study. %and texts. Since it's impossible to design icons for each individual landmark, we divided landmarks into different categories based on PLV's preferences from the formative study. 
Section \ref{Landmark selection} details the landmark icon selections. Besides the icons, we also used text to describe the specific landmark.

%TC:ignore
%TC:endignore

\subsubsection{In-situ Labels}
We positioned large virtual in-situ icons and text in vibrant colors at the landmark's physical position to augment each individual landmark (\textit{DG} 4), as shown in Figure \ref{fig:VisiMark}C. We opted for icons and text instead of outlines and color coating to align with the icon and text labels displayed on the Signboards. The size and font color of the in-situ labels can be customized based on PLV's preferences. Moreover, to enable PLV to easily see the augmentations from different angles, the labels rotated automatically based on the user's position to ensure facing the user all the time (\textit{DG 3}). %We anticipated that these designs would enhance landmark visibility as potential global landmarks, addressing challenges related to visual acuity and contrast. 
Additionally, we numbered the same types of landmarks if they appeared multiple times in the environment to enable PLV to better distinguish them.

\subsubsection{Landmark and Icon Selection}\label{Landmark selection}
%To answer our RQ 2, one of our goals is to investigate what types of landmarks PLV would prefer to augment. Thus, 
VisiMark supported all landmark categories used by PLV and provided icons as augmentations. We categorized landmarks into five groups according to PLV's landmark preferences. Specifically, based on the original visual, cognitive, and structural landmark categories \cite{sorrows1999nature}, we expand the cognitive landmarks into \textit{Information}, \textit{Accessibility}, and \textit{Emergency} landmarks since they are the most important landmarks preferred by PLV for safety and accessibility purposes. We did not further unfold visual and structural landmarks since visual landmarks can be seen by PLV with their residual vision and structural landmarks (e.g., space structure) were mostly presented by the Signboard. %We designed icons for each type of landmarks as augmentations. %In VisiMark, we selected augmented landmarks according to these categories, while recognizing that there are many other potential landmarks along the routes. 
%For visual landmarks, since participants can effectively identify obvious features like changes in lighting and area "silhouettes", we did not further subdivide this category. For structural landmarks, we focused on specific individual landmarks and represented the overall layout in the signboard design. For cognitive landmarks, we further divided them into information, accessibility and emergency, as PLV are particularly attentive to landmarks with safety and accessibility implications that can aid or impede their mobility. 
We elaborate the five types of landmarks and their corresponding icons: 
%\yuhang{this added sentence is too arbitrary. You may want to add the rationale on the higher level, e.g., we investigate what landmarks user prefer to augment, our current design support all types of landmarks commonly used by PLV. We categorized the landmarks into xxx ...sometimes i added a comment here does not mean your edits should be exactly at this place. You want to think about your paper as whole and make adjustments in context.---my comments to you is not a small task in the local area. It's a start for you to think what is the potential issue with the current draft and how to address it.}
\begin{itemize}
    \item \textit{Visual \includegraphics[width=0.3cm]{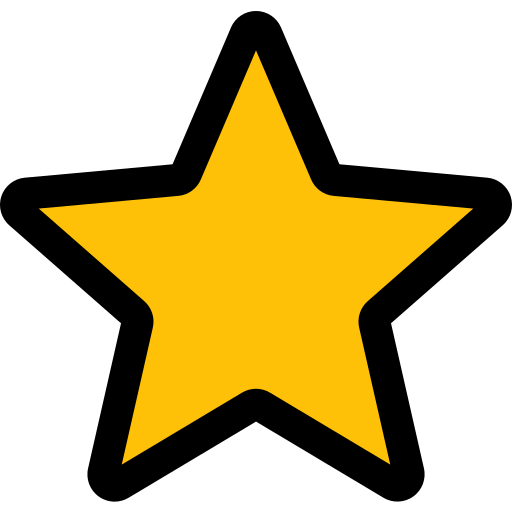}}: visually obvious (e.g., high contrast, large or bright) landmarks that may draw attention, such as large green double doors and a red cork board.
    \item \textit{Information \includegraphics[width=0.3cm]{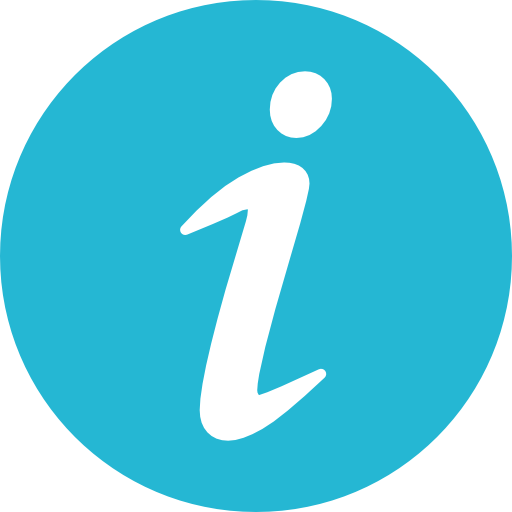}}: landmarks that convey important information or support certain functions, such as restrooms or laboratories. 
    \item \textit{Accessibility \includegraphics[width=0.3cm]{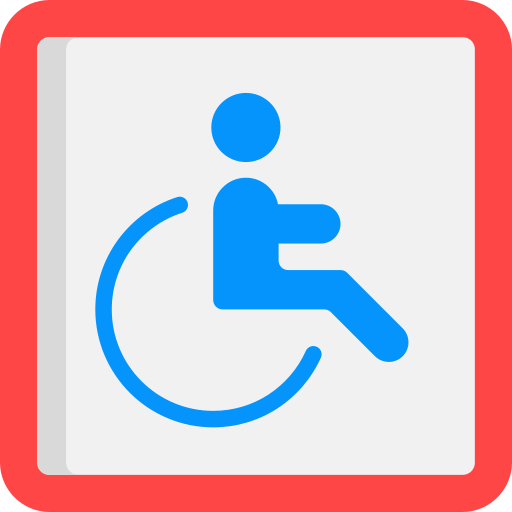}}: landmarks that facilitate accessibility and mobility, such as elevators, ramps, and railings.
    \item \textit{Emergency \includegraphics[width=0.3cm]{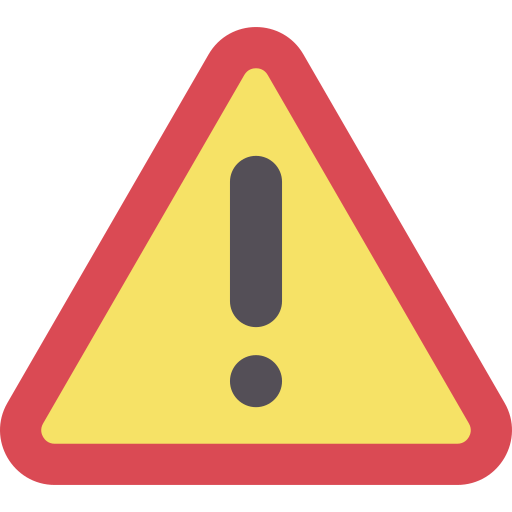}}: landmarks that indicate potential dangers or emergency resources, such as biohazard danger signs and automated external defibrillator (AED).
    \item \textit{Structural \includegraphics[width=0.3cm]{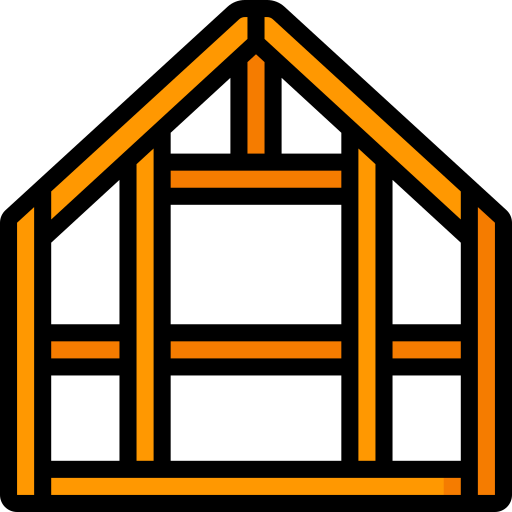}}: landmarks that represent structural elements, such as windows and stairs. They mark the building's architectural features.
\end{itemize}

\subsection{Prototype Implementation}
\colorchange{
We built our VisiMark prototype on Microsoft HoloLens 2. We pre-scanned the environment with HoloLens using the spatial mapping API \cite{microsoft-spatial-mapping} and added a QR code \cite{microsoft-qr-code-tracking} at the beginning of each route to align the landmark augmentation interfaces with the route. Since our research focused on landmark augmentation designs and their impact, we adopted this pre-scanning method to ensure accurate alignment and avoid potential confounding effects caused by real-time route and landmark recognition algorithms (e.g., recognition errors and delays). After the alignment, the system relied on the integrated motion and position tracking capability of HoloLens to track the user's position along the routes and generate corresponding augmentations at certain intersections and landmarks. 
%to avoid the confounding effect of potential computer vi added a QR code Due to the similarity of indoor environments, we chose to scan QR codes \cite{microsoft-qr-code-tracking} at the starting point of each route to hard-code landmark locations instead of spatial anchors \cite{microsoft-spatial-anchors}. 
We built the prototype with Unity 2021.3.31f1 and Microsoft Mixed Reality Toolkit 3 \cite{microsoft-mrtk3-overview}.}

\section{EVALUATION METHOD} \label{sec:eval}
We seek to evaluate the effectiveness and impact of VisiMark on PLV's landmark perception. Specifically, our study will answer: (1) How does VisiMark affect PLV's retracing performance? (2) How does VisiMark affect PLV's mental model development and landmark selections? (3) What types of landmarks do PLV need to augment, and (4) what are the suitable landmark augmentations preferred by PLV? %We collected both quantitative performance data (e.g., mental map accuracy, route retracing time and accuracy) and qualitative feedback from participants and brainstormed with them on potential landmark augmentation designs.% ideas on how to augment them. 
% \yuhang{might be good to list specific research question for the study: (1) How does VisiMark affect PLV's retracing performance? (2) How does VisiMark affect PLV's mental modal development? (3) How does VisiMark affect PLV's landmark selections? (4) What landmarks PLV prefer to augment and how to augment them? }

\subsection{Participants}
We recruited 16 low vision participants (8 females, 7 males, 1 non-binary) whose ages ranged from 18 to 72 ($Mean = 44.8$, $SD = 19.8$). Five participants had prior AR experiences. Two participants (T8, T11) were legally blind but had functional vision to navigate. All participants wore eye glasses to partially correct their acuity. The recruitment method and compensation were the same as the formative study. Table \ref{tab:demographics_evaluation} shows participants' demographic information in detail. Two participants also took part in the formative study: T2 (P5) and T10 (P3).

%TC:ignore
\begin{table*}[ht]
\centering
\scriptsize
\begin{tabular}{>{\centering\arraybackslash}m{0.5cm} >{\centering\arraybackslash}m{0.7cm} >{\centering\arraybackslash}m{2.5cm} > {\centering\arraybackslash}m{1cm} >{\centering\arraybackslash}m{2cm} >{\centering\arraybackslash}m{3cm}  >{\centering\arraybackslash}m{4cm} > {\centering\arraybackslash}m{1.2cm}}
\Xhline{2\arrayrulewidth}
\textbf{ID} & \textbf{Age/ Gender} & \textbf{Diagnosis} & \textbf{Legally Blind} & \textbf{Visual Acuity} & \textbf{Field of View (FOV)} & \textbf{Other Visual Difficulties} & \textbf{Prior AR Experiences} \\
\Xhline{2\arrayrulewidth}
T1 & 66/M & Right eye damaged & N & R: 20/60; L:20/20 & Two-thirds of the lower vision loss in right eye & Less sensitive to light; experienced discomfort from bright side glare, causing vision flare; depth perception issue &  Y\\
\hline
T2 & 72/M & Macular degeneration & N & 20/40 \& 20/50 & Full & N/A& N \\
\hline
T3 & 31/F & Retinitis pigmentosa & N & Not know & Like a donut, the central and peripheral are blurry & Sensitive to light and contrast & N \\
\hline
T4 & 55/M & A complete loss of vision in left eye & N & Near sighted right eye, blind left eye & Full & No depth perception & Y \\
\hline
T5 & 19/Non-binary & Blind left eye& N& Near sighted  right eye, blind left eye & Full & No depth perception & Y \\
\hline
T6 & 69/M & Retinopathy of prematurity & N & R: 20/400; L: 20/30 & Central vision loss in right eye & Sensitive to light in right eye; depth perception issue & N \\
\hline
T7 & 26/F & Atrophy of optic nerve & N & 20/400 \& 20/70 & Central is a lot worse than peripheral & Sensitive to contrast & N \\
\hline
T8 & 40/M & Macular myopic degeneration & Y & R: 20/2400; L: 20/30 (with correction) & Central vision degeneration in both eyes, peripheral vision loss in right eye & Sensitive to light; color blind, confused between green, yellow, red and orange; depth perception issue & N \\
\hline
T9 & 50/F & Blind right eye & N & R: <20/200; L: 20/20 (with correction) & Full & Sensitive to light; depth perception issue & Y \\
\hline
T10 & 71/F & Chemo induced type of neurological cornea issue and macular degeneration & N & R: 20/100; L: 20/200 & Full & Photophobia; confused between red and blue, as well as red and black & N \\
\hline
T11 & 58/F & Retina pigmentosa & Y & 20/50 \& 20/50 & Peripheral vision loss & Sensitive to light; confused between pink and orange, as well as brown, blue and grey; depth perception issue & N \\
\hline
T12 & 27/F & Weaker eye muscle and double vision & N & R: 20/800; L: 20/400 (with correction: R: 20/25; L: 20/30) & Full & Depth perception issue & N \\
\hline
T13 & 59/M & Glaucoma & N & 20/2400 in one eye, relativtly good acuity in the other & Top left vision loss in right eye and top right vision loss and peripheral vision loss in left eye & N/A & N \\
\hline
T14 & 33/F & Congenital cataract and amblyopia in right eye & N & R: 20/500; L: 20/20 & 70 degrees & No depth perception & N \\
\hline
T15 & 22/M & Blurred vision that cannot be corrected by lens or contact lens & N & R: 20/40 (with correction); L: 20/40 (with correction) & Full & Depth perception issue when it gets darker & Y \\
\hline
T16 & 18/F & Ocular albinism & N & R: 20/100; L: 20/80 & Full & Sensitive to light; depth perception issue & N \\
\Xhline{2\arrayrulewidth}
\end{tabular}
\caption{Participant demographic information for the final evaluation.}
\label{tab:demographics_evaluation}
\end{table*}
%TC:endignore

\subsection{Apparatus}
% To support the study, we implemented the VisiMark interface on Microsoft HoloLens 2 and prepared four similar indoor routes for navigation tasks. We describe our environment setup and system implementation details below.

\begin{figure*}[ht]
    \centering
    \begin{subfigure}{\textwidth}
        \centering
        \includegraphics[width=0.9\textwidth]{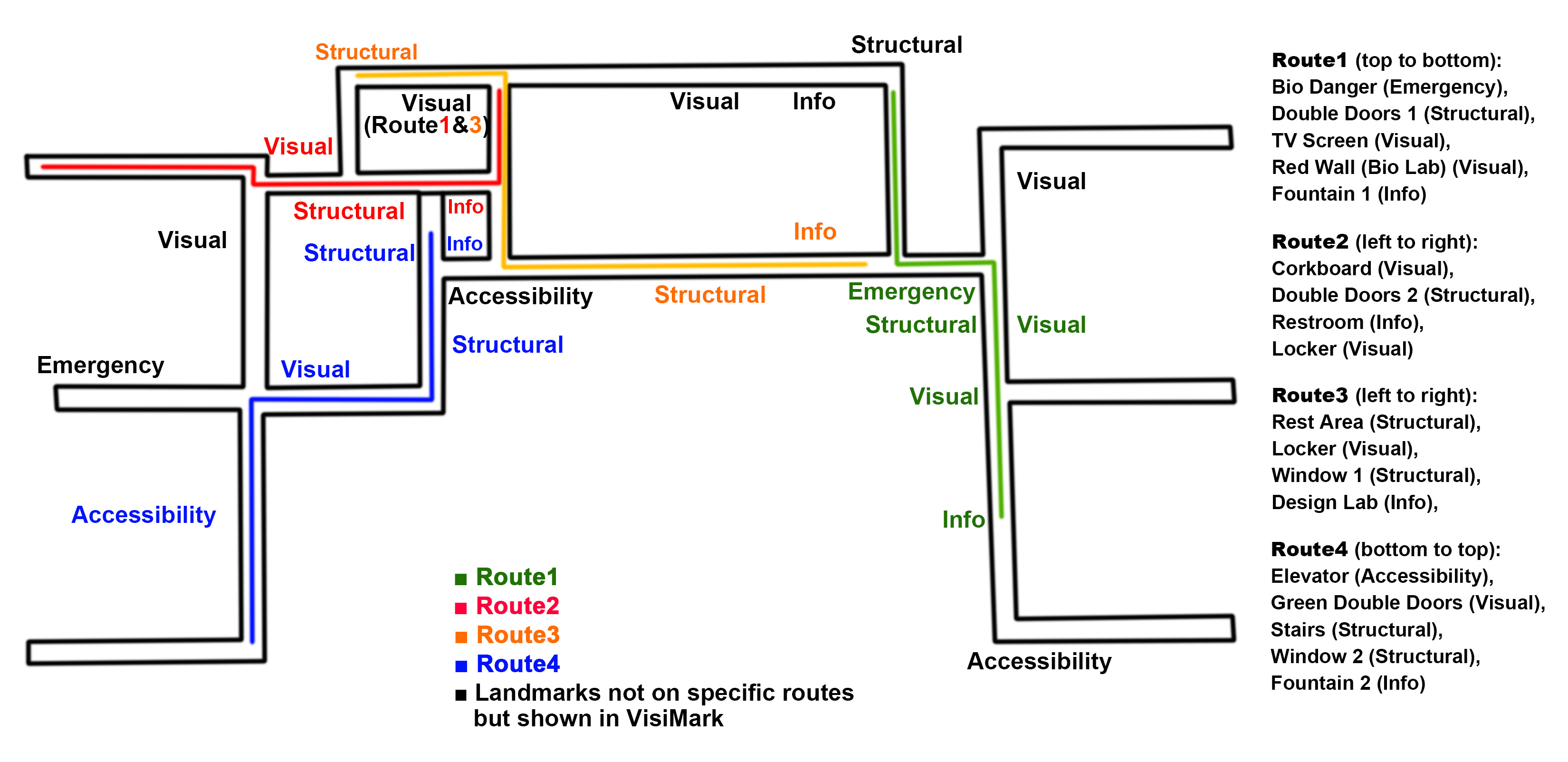}
    \end{subfigure}
    \caption{Four routes in the study environment with labeled landmarks augmented in VisiMark, including both those along the routes and other landmarks in the space displayed on Signboards. Detailed landmarks and their types for each route are listed on the right.}
    % \yuhang{add a table on the side of the map, labeling the landmarks and types for each route maybe following left to right and up to down order.}
    \Description{This image shows four routes in the study environment with labeled landmarks augmented in VisiMark, including both those along the routes and other landmarks in the space displayed on Signboards. Detailed landmarks and their types for each route are listed on the right.}
    \label{fig:route_planning}
\end{figure*}

% \textbf{\textit{Environment Setup.}} 
We conducted the study in a well-lit indoor environment. We used one floor in one of our academic buildings as our study environment and planned four similar routes with similar lengths and equal number of decision points \cite{ishikawa2008wayfinding, may2020spotlights}. All routes were approximately 65 meters long and included three decision points, consisting of two turns and one additional decision point, as shown in Figure \ref{fig:route_planning}. We also augmented similar number of landmarks on each route (4-5 landmarks per route). Our landmark selection ensured the coverage of all five categories of landmarks, with each route covering 3-4 landmark categories. We chose landmarks with the most salient visual, informative, accessible, emergent, or structural features within each category, such as a large TV screen as visual landmarks and an elevator as accessibility landmarks, as shown in Figure \ref{fig:route_planning}. %\yuhang{i'm still not highly convinced...we have five categories, why here you only mentioned three types of features?.}%In particular, due to the limitations of the indoor environment, we counted two consecutive turns in Route 2 as a single turn. 
We also set up a training area with one intersection and three hallways on another floor for the tutorial session.
% Specifically, we selected four landmarks on Route 2 and Route 3 to augment, and five landmarks on Route 1 and Route 4.

%TC:ignore
%\vspace{-1ex}
%\vspace{-1ex}
%TC:endignore

% \textbf{\textit{Implementation.}} 
\colorchange{Since PLV may experience different visual conditions, we allowed participants to customize the AR interface to refine their visual experience and ensure the visibility of landmark augmentations.} Current customization included toggling on and off each design component (e.g., icon and text labels on Signboards, color-coded hallways) and adjusting parameters of each component (e.g., font color, font size, icon size, Signboard size). \colorchange{We implemented the customization through the method of Wizard of Oz \cite{dahlback1993wizard}, where the participants used speech commands to customize the interface and the experimenter made the adjustments accordingly via a smartphone application we developed.} The phone application also tracked the system data of VisiMark (e.g., virtual elements shown on the AR display) and \colorchange{presented it to the experimenter via text logs to allow them to better understand the participant's experience}. The phone application was built with React Native and communicated with the HoloLens via WebSocket. 

% and log system data  in real-time during the navigation to ensure the experiment went smoothly. 

% To enable customization, we employed Wizard of Oz studies , allowing the experimenter to tailor VisiMark according to the participants' specific needs.

\subsection{Procedure}
The study consisted of a single session that lasted approximately two hours. We started with an initial interview, asking about participants' demographic information, visual conditions, and experiences with AR and navigation. We also asked about their experiences with landmarks, such as what landmarks they commonly use and why.
%Detailed questions are listed in Appendix \ref{}.

We then conducted a tutorial to familiarize participants with our system. We introduced each design feature of the system and explained the meaning of different icons for each landmark category. Participants were then able to customize some design parameters based on their visual ability, including the size of the Signboard, the size of icons (both on Signboard and In-situ Labels), and the size and color of the text labels. Additionally, the height of the Signboards was adjusted based on the participants' heights for an eye-level rendering. During the tutorial, participants were encouraged to walk around the tutorial area to freely experience VisiMark.

% Our aim was to address AR learning challenges effectively through the tutorial.

After feeling fully familiar and comfortable with VisiMark, each participant performed four trials of navigation tasks under two conditions (two trials per condition): (1) wearing an AR device and (2) not wearing an AR device. Each navigation task consisted of three phases: \textit{Exploration}, \textit{Mental Map Drawing}, and \textit{Retracing}. In the exploration phase, participants freely navigated along a pre-planned route by following verbal turn-by-turn instructions. \colorchange{A researcher on the team followed the participant and provided verbal instructions when the participant approached each intersection (e.g., ``Turn right at this intersection''). We provided verbal guidance (1) to simulate users' real-world indoor navigation experiences since low vision people often ask others for verbal route descriptions during navigation \cite{szpiro2016finding}, and (2) to avoid potential interference with the visual landmark augmentations provided by VisiMark.} During this navigation, participants were free to look around, observe the environment, and walk at their comfortable pace. After reaching the destination, participants were asked to draw a mental map of the route, including any landmarks they remembered along the way, as accurately as possible (mental map drawing phase). Participants also assessed their confidence in the map's accuracy with a rating between 1 to 7, where 7 meant extremely confident and 1 meant extremely unconfident. Finally, in the retracing phase, we led the participants back to the starting point along a different route to prevent them from seeing the original path. Participants were then asked to retrace the original path and reach the destination as quickly as possible without any verbal instructions.

We counterbalanced the order of the two conditions (wearing or not wearing the AR device) and the four routes using Latin Square, resulting in eight (two conditions $\times$ four route orders) combinations. With 16 participants recruited in the study, each combination was assigned to two participants randomly.% assigned to the conditions, with each condition being repeated twice. %The detailed arrangement is illustrated in Table \ref{tab:latin_square}.

% \begin{table}[ht]
% \centering
% \caption{The Latin Square used to counterbalance the four routes and the two conditions (wearing or not wearing the AR device).}
% \label{tab:latin_square}
% %\scriptsize
% \begin{tabular}{c c c c c}
% \Xhline{2\arrayrulewidth}
% 1 & Route 1 \& with AR & Route 2 \& without AR & Route 3 \& with AR & Route 4 \& without AR \\
% \hline
% 2 & Route 2 \& without AR & Route 3 \& with AR & Route 4 \& without AR & Route 1 \& with AR \\
% \hline
% 3 & Route 3 \& with AR & Route 4 \& without AR & Route 1 \& with AR & Route 2 \& without AR \\
% \hline
% 4 & Route 4 \& without AR & Route 1 \& with AR & Route 2 \& without AR & Route 3 \& with AR \\
% \hline
% 5 & Route 1 \& without AR & Route 2 \& with AR & Route 3 \& without AR & Route 4 \& with AR \\
% \hline
% 6 & Route 2 \& with AR & Route 3 \& without AR & Route 4 \& with AR & Route 1 \& without AR \\
% \hline
% 7 & Route 3 \& without AR & Route 4 \& with AR & Route 1 \& without AR & Route 2 \& with AR \\
% \hline
% 8 & Route 4 \& with AR & Route 1 \& without AR & Route 2 \& with AR & Route 3 \& without AR \\
% \Xhline{2\arrayrulewidth}
% \end{tabular}
% \end{table}

\subsubsection{Exit Interview}
We ended our study with a semi-structured interview, discussing participants’ landmark selections without VisiMark, the impact of our system on their landmark selections, and their suggestions for landmark augmentations in AR. Participants were asked to rate their perceived effectiveness and comfort for completing the tasks with and without the system, as well as the distraction and learnability of VisiMark on a 7-point Likert scale. \colorchange{Detailed interview questions can be found in Appendix \ref{Interview Questions for the Evaluation}.}%Participants were asked to rate their perceived effectiveness and comfort for completing the tasks with and without the system on a scale from 1 to 7, where 7 meant extremely effective/comfortable and 1 meant extremely ineffective/uncomfortable. Participants also assessed the distraction and learnability of VisiMark on a scale from 1 to 7, where 1 indicated least distracting/easy to learn and 7 indicated most distracting/easy to learn.

% We collected participants' subjective feedback evaluating the effectiveness and comfort ratings for completing the task both with and without the system. Additionally, we gathered Likert scale ratings to evaluate the distraction and learnability of the system. 
% Detailed questions are listed in Appendix \ref{}.

\subsection{Analysis}
We collected both quantitative measures and quantitative feedback from the study. Our analysis methods are as below.

\subsubsection{Measures \& Statistical Analysis}

% \paragraph{Wayfinding task evaluation}
We defined two measures for the navigation task: (1) \textit{Retracing Time}, the time spent by a participant to retrace their route from the starting point to the destination; and (2) \textit{Retracing Correctness}, a binary value to indicate whether a participant made a wrong turn or failed to reach the destination.

% \paragraph{Mental map evaluation}
We also defined four measures for mental map evaluation: (1) \textit{Correctness of Turns}, the number of correct turns minus the number of incorrect extra turns; (2) \textit{Correctness of the Route Segment Lengths}, the number of adjacent segments that have the correct relative length relationship (i.e., longer than or shorter than), following the method in Aginsky et al. \cite{aginsky1997two}; (3) \textit{Total Landmark Recall}, the number of landmarks that participants recalled and actually exist along the route, including landmarks that were augmented by VisiMark and those that were not augmented but still remembered as landmarks by the participants; and (4) \textit{Recall of Augmented Landmarks}, the ratio of augmented landmarks recalled by the participant correctly to the total number of augmented landmarks on each route. For the trials where participants \textit{did not} wear VisiMark, we calculated this measure by assessing the recall of ``VisiMark landmarks'' (i.e., the specific landmarks that VisiMark would augment on the route). By comparing this measure between the with and without VisiMark conditions, we were able to evaluate the overlap between VisiMark landmarks and the landmarks naturally selected by PLV, thus deriving the impact of VisiMark on PLV's landmark selections.%\yuhang{i'm wondering whether we should do a between-subject comparison for each route (even not statistically), since the routes differ so much. Not sure it makes sense to compare between different routes via within subject method.}
% I calculate ANOVA for route on Recall of Augmented Landmarks and there is no significant effect

For all the measures, we had a within-subjects factor named \textit{Condition} with two levels: with VisiMark and without VisiMark. To validate the counterbalancing, we involved another within-subject factor \textit{Order}, and found no significant effect of \textit{Order} on any measures. We checked the normality of each measure using Shapiro-Wilk test. None of these measures were normally distributed, so we used the Aligned Rank Transform (ART) ANOVA to evaluate the effect of Condition on different measures. We used partial eta squared ($\eta ^2_p$) to calculate the effect size, with 0.01, 0.06, 0.14 representing the thresholds of small, medium and large effects \cite{cohen2013statistical,wang2024gazeprompt}.

\subsubsection{Qualitative Analysis}
We audio-recorded all studies and transcribed interviews using an automatic transcription service. Similar to the formative study, we analyzed the transcripts using thematic analysis. Two researchers open-coded five identical samples (31\% of the data) independently and developed an initial codebook by discussing their codes to resolve any disagreements. The two researchers then separated the remaining transcripts and coded independently based on the initial codebook. During this process, we periodically checked each other's work and discussed the codes to ensure consistency. After the researchers reached an agreement, new codes were added to the codebook.

We developed themes from the codes using a combination of inductive and deductive approaches \cite{braun2006using}. Our research aimed to evaluate our landmark-based navigation system, with a specific focus on what kinds of landmarks should be augmented for PLV and what are suitable augmentations for landmarks. Our high-level theme generation was thus guided by these objectives, following the deductive approach. Within each theme, we employed the inductive approach, generating sub-themes by clustering relevant codes using axial coding and affinity diagrams. Once initial themes and sub-themes were identified, researchers cross-referenced the original data, the codebook, and the themes to make final adjustments, ensuring all codes were correctly categorized. Our analysis resulted in 6 themes with 17 sub-themes, detailed in Appendix \ref{Codebook for Evaluation}.%\yuhang{would be good to add the specific theme and subtheme numbers for the final evaluation.}
\section{EVALUATION RESULTS}
We report the effectiveness and impact of VisiMark via quantitative and qualitative data, \colorchange{as summarized in Table \ref{tab:quan_measures}}. Additionally, we developed a taxonomy for landmarks to be augmented for PLV and explored the design space for augmenting these landmarks.
%\colorchange{We summarize VisiMark’s impact on user performance and experience based on quantitative measures in Table \ref{tab:quan_measures}, and landmark taxonomies for PLV in different contexts in Table \ref{tab:landmark_taxonomies}.}%\yuhang{the findings are somehow structured in a quite arbitrary way, with system evaluation that includes all kinds of things... break them down based on research questions or themes.}

\subsection{Effect of Landmark Augmentations on Route Retracing}
We first evaluated VisiMark through participants' performance in route retracing. %We also gathered their comments on different components and aspects of the system, such as distraction and learnability. Our findings are elaborated below.
Although there was no significant effect of Condition on \textit{Retracing Time} ($F_{1,47} = 0.001$, $p = .975$, $\eta _p^2 < 0.01$), participants finished the retracing task slightly faster with VisiMark ($Mean_{with}=58.3s$, $SD_{with}=16.9s$, $Mean_{without}=59.7s$, $SD_{without}=20.7s$). As for the correctness of retracing, we found no significant effect of Condition on \textit{Retracing Correctness} ($F_{1,47} = 0.370$, $p = .546$, $\eta _p^2 < 0.01$). However, four participants made errors when they were wearing the system, while eight participants made errors without the system during retracing. The data showed that participants performed slightly better in retracing with VisiMark. Moreover, participants (10/16) mentioned explicitly that they felt themselves navigating more effectively with VisiMark since the augmentations provided more guidance along the retracing. As T8 explained, ``It was definitely easier [when using VisiMark] because I had that visual with landmarks. And I was able to, like, I almost bypassed the first turn, but then I saw the [augmented] locker or whatever it was, [I know] oh yeah, I gotta turn to [that way].'' Some participants (T7, T11, T12) memorized the turns based on the lengths of hallways provided by Signboards, as T11 said, ``I remember just going to [the] short[er] arrow way. That's how I remembered which way to go.''

\subsection{Effect of Landmark Augmentations on Mental Map Development}
We evaluated participants' mental maps from route and landmark perspectives. % via different aspects, including turns, route segments, and landmarks. % including \textit{Correctness of Turns}, \textit{Correctness of the Route Segment Lengths}, \textit{Landmark Recall Count} and \textit{Recall of
%Augmented Landmarks}, along with their qualitative feedback. 
Our findings indicate that VisiMark helped participants better perceive hallway structures, increased their focus on landmarks, and led to a shift in landmark selections. We elaborate on these findings below.

%TC:ignore
\begin{figure*}[ht]
    \centering
    \begin{subfigure}{0.9\textwidth}
        \centering
        \includegraphics[width=0.9\textwidth]{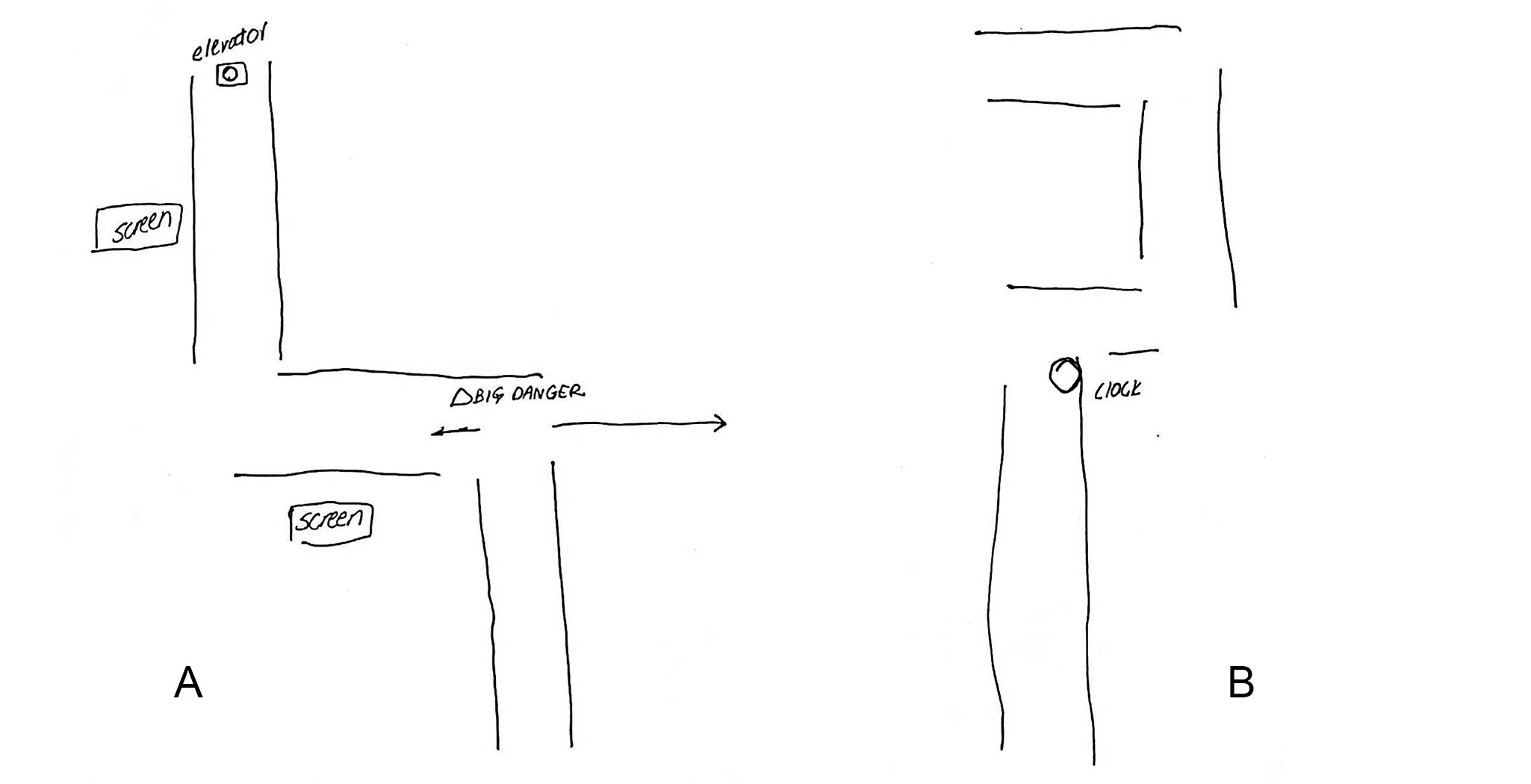}
    \end{subfigure}
    \caption{Examples of participants' mental maps: (A) T7's mental map of route 1 with VisiMark, and (B) T7's mental map of route 2 without VisiMark. We observed a landmark selection shift here, as T7 chose more meaningful cognitive landmarks with VisiMark.}
    \Description{
    This image shows examples of participants' mental maps: (A) T7's mental map of route 1 with VisiMark, and (B) T7's mental map of route 2 without VisiMark. We observed a landmark selection shift here, as T7 chose more meaningful cognitive landmarks with VisiMark.
    }
    \label{fig:mental_maps}
\end{figure*}
%TC:endignore

\subsubsection{Perceived Hallway Structures.} We found no significant difference in \textit{Correctness of Turns} ($F_{1,47} = 2.879$, $p = .096$, $\eta _p^2 =0.06$) and \textit{Correctness of the Route Segment Lengths} ($F_{1,47} = 0.058$, $p = .811$, $\eta _p^2 < 0.01$) in participants' mental maps between with and without VisiMark. %^Participants drew the route map with a high degree of accuracy in most cases, with 87\% of turns and 89\% of route segments in 64 trials containing no more than one error.\yuhang{is this because they all painted them correctly? if so, we should say that-- Unfortunately not...} 
However, five participants reported that they could better perceive hallway structures using VisiMark than without. For example, T7, T11, and T12 memorized the lengths of the hallways due to the virtual representation of the hallways on Signboards and leveraged this information in their mental maps. As T7 said when she drew the mental map, ``I remember the big screen sign and then see this was [a] shorter hallway... and I think [another hallway] was [a] longer hallway.'' %\yuhang{add a quote about the "turning into the shorter hallway."---also, maybe this should go to the last section since it's about retracing?}
%helped them learn the environment and build more accurate mental maps, as the
%(T2,T7,T11,T12,T14)%and the corresponding hallway lengths, noting that they should turn into the shorter hallway during the retracing. 

\subsubsection{Increased Focus on Landmarks.} There was no significant effect of Condition on \textit{Total Landmark Recall} ($F_{1,47} = 0.556$, $p = .460$, $\eta _p^2 =0.01$). However, all participants thought the system increased their attention to landmarks, which helped them notice (12/16), identify (4/16), locate (10/16), and memorize landmarks (8/16). T6 and T11 both mentioned that VisiMark helped with memorization because it made the information more comprehensible, as T6 explained, ``When you make the [landmarks] more notable, it makes the association and your brain better.'' 

Moreover, three participants (T7, T13, T16) noted that they relied more on landmarks in navigation than before, considering this a positive change as it helped them memorize the environment better, otherwise they only cared about turns and the destination. As explained by T7, ``When I was not using the system, I was kind of memorizing more like, I go right, and then left and then right. But then when I was using the system... I like remember the arrows appearing and their length.'' T16 added, ``But once [the system] was taken off, it can be hard to remember where things were... [Without the system] I only know where I need to go and not like things around it.'' % \yuhang{this is interesting but not sure it should go under "Total landmark recall." Maybe we should rename this section}

\subsubsection{Impact on Landmark Selection.}
\label{Impact on Landmark Selection}
We found a significant difference of Condition on \textit{Recall of Augmented Landmarks} ($F_{1,47} = 22.428$, $p < .001$, $\eta _p^2 =0.32$), which indicates the change in landmark selections caused by VisiMark. Nine participants noted that VisiMark enabled them to notice landmarks that they had previously overlooked. %with T15 mentioning increased attention to visually salient landmarks. 
For example, five participants included the augmented bio-danger sign in their mental map of Route 1 with VisiMark, whereas none noticed it without the system since they were small. %Similarly, three out of eight participants painted a visually obvious TV screen with VisiMark, while none did without it. 
The system also enabled participants to notice objects in their blind spots. T4, who was blind in the left eye, noticed an elevator on his left with VisiMark on Route 4, while T5, also blind in the left eye, missed it on the same route without VisiMark.%The elevator was mentioned as a landmark 8 times with VisiMark, compared to only 3 times without it. \yuhang{this is very important but sort of buried at the end of this paragraph. "Facilitating landmark selection" is an important theme. You mentioned it in last section, and a little bit here. We should merge them and separate them into its own theme, with some quantitative comparison between the types of landmarks people recall that are augmented vs not augmented.} 

Landmark augmentations enabled participants (11/16) to locate more functional facilities (e.g., elevators, restrooms) and unique objects (e.g., distinctive double doors that were not highly contrasting with the environment), rather than just visually salient landmarks (as shown in Figure \ref{fig:mental_maps}), by providing additional semantic information for mental model development and memorization (7/16). As T4 remarked, ``For remembering the route, having the landmarks stated and mentioned [with] an extra [cue], that's a good point. Because the text did help, [like] an icon for a green [double] door, I would not have remembered the double door [without VisiMark].'' 
% \yuhang{this quote does not really match the argument before. The quote is talking about the overview, but the argument is for suitable landmark selections. need to replace the quote.} 
Some participants (T2, T3, T7, T11) started to pay attention to once-augmented landmarks even without wearing the system. They considered this beneficial, as the system helped them identify important landmarks more effectively. As T7 commented, ``The arrows (Signboards) really helped my understanding of the route because I wasn't like, looking at weird things on the ceiling.'' She added, ``Most of the objects that I was picking out when I did the routes with the glasses were things that you guys had in your system, rather than just ordinary things in the ordinary environment.'' Although T10 noted that the augmentation conflicted with her own way of identifying landmarks, as the system reinforced her to notice some cognitive landmarks that were not visually salient to her, she was positive about the perception change and adjusting to the system. This subtle shift in landmark selection, facilitated by the system, helped participants pick up more memorable and meaningful landmarks.
% \yuhang{these are about landmark selection again. We should really merge qualitative feedback that conveys the same theme, instead of repeating them in different subsection. It is important to find qualitative feedback to support quantitative data, but they should match.}
% Additionally, most participants (15/16) mentioned that the landmark augmentations enabled them to locate more functional facilities (11/16) as opposed to landmarks that are only visually salient, thus offering more semantic information to learn and memorize (7/16). 
% As T8 said, ``[VisiMark] gives you like a mental preparation ... like in a building that I don't know, I'm not familiar with. So it's it is helpful.'' 

%TC:ignore
\begin{table*}[h]
\centering
\small
\begin{tabular}{>{\raggedright\arraybackslash\color{brownishred}}m{3cm} >{\raggedright\arraybackslash\color{brownishred}}m{3cm} >{\raggedright\arraybackslash\color{brownishred}}m{1.7cm}
>{\arraybackslash\arraybackslash\color{brownishred}}m{8.5cm}}
\Xhline{2\arrayrulewidth}
\textbf{Tasks} & \textbf{Measures} & \textbf{Statistical Results} & \textbf{Result Interpretation \& Key Qualitative Findings} \\
%& \textbf{Details}\\
\Xhline{2\arrayrulewidth}
\multirow{4}{*}{Retracing Performance}& Retracing Time&$p=.975$&Participants retraced slightly faster with VisiMark, although the difference was not significant.\\
%&$F_{1,47} = 0.001$, $\eta _p^2<0.01$\\
\cline{2-4}
& Retracing Correctness& $p=.546$&Fewer participants made retracing errors with VisiMark, although the difference was not significant.\\
%&$F_{1,47} = 0.370$, $\eta _p^2 <0.01$\\
\hline
\multirow{8}{*}{Mental Map Development}& Correctness of Turns& $p=.096$&While no significant differences were observed in the mental\\
%&$F_{1,47} = 2.879$, $\eta _p^2 =0.06$\\
\cline{2-3}
& Correctness of the Route Segment Lengths&$p=.811$& map development, five participants reported that they could better perceive hallway structures with VisiMark.\\
%&$F_{1,47} = 0.058$, $\eta _p^2 <0.01$\\
\cline{2-4}
\multirow{3}{*}{}& Total Landmark Recall& $p=.460$&While no significant differences were found, all participants reported that VisiMark increased their attention to landmarks.\\
%&$F_{1,47} = 0.556$, $\eta _p^2 =0.01$\\
\cline{2-4}
& Recall of Augmented Landmarks& $p<.001$***&VisiMark significantly changed participants' landmark selection from only visually-salient objects to cognitive landmarks that are more important and meaningful.\\
%&$F_{1,47} = 22.428$, $\eta _p^2 =0.32$\\
\hline
\multirow{8}{*}{Subjective Experiences}& Effectiveness & $p=.036$*& Participants perceived VisiMark to be significantly more effective in supporting indoor navigation than the baseline.\\
%&$Mean_{with}=5.41$, $SD_{with}=1.02$, $Mean_{without}=4.41$, $SD_{without}=1.64$\\
\cline{2-4}
\multirow{5}{*}{}& Comfort Level& $p=.037$*&VisiMark significantly improved participants' comfort level in indoor navigation compared to the baseline condition.\\
%&$Mean_{with}=5.78$, $SD_{with}=1.05$, $Mean_{without}=4.72$, $SD_{without}=1.71$\\
\cline{2-4}
& Learnability&$Mean=6.09$, $SD=0.78$&VisiMark was easy to understand and learn, with 14 participants rating its learnability at six or higher. \\
\cline{2-4}
& Distraction&$Mean=3.03$, $SD=1.62$&
VisiMark was not distracting, with 11 participants rating its distraction level at three or below.\\
\Xhline{2\arrayrulewidth}
\end{tabular}
\caption{\colorchange{Impact of VisiMark on user performance and experiences. Statistical significance is noted as follows: * for $p<0.05$, ** for $p<0.01$, and *** for $p<0.001$.}}
\label{tab:quan_measures}
\end{table*}

%TC:endignore

\subsection{Experiences with VisiMark}
\subsubsection{Effectiveness \& Comfort}
\label{Effectiveness and comfort}
We employed a paired t-test to assess the participants' perceived effectiveness and comfort level with VisiMark in navigation. Our analysis revealed a significant difference in effectiveness ratings between conditions with and without the system ($t_{15} = 2.30$, $p = .036$), suggesting that VisiMark significantly improved perceived effectiveness ($Mean_{with}=5.41$, $SD_{with}=1.02$, $Mean_{without}=4.41$, $SD_{without}=1.64$). Similarly, we observed a significant difference in comfort ratings between with and without the system ($t_{15} = 2.30$, $p = .037$), suggesting that VisiMark significantly improved task comfort ($Mean_{with}=5.78$, $SD_{with}=1.05$, $Mean_{without}=4.72$, $SD_{without}=1.71$). Most participants (13/16) found VisiMark to be mentally comfortable, while T9, T10 and T14 reported they were more comfortable without the system because of the learning curve. As T14 explained, ``It's just what I'm used to versus wearing the [AR glasses].'' Four participants noted that discomfort mainly came from the AR device; T10 mentioned that the device conflicted with her own glasses. Furthermore, two participants (T5, T13) said that they were uncomfortable using the AR device in public, as they felt it made them more noticeable to passersby.
% \yuhang{why? privacy? awkward looking?}.

\subsubsection{Learnability}
Participants highly rated the learnability of VisiMark with a mean of 6.09 ($SD=0.78$) on a 7-point Likert scale, indicating the system was very easy to understand and use. 
% T6 and T8 noted the texts were simple to understand.
However, most participants (13/16) mentioned that there was a learning curve in getting used to VisiMark while the learning curve was short (5/16). Four participants (T2, T4, T8, T10) mentioned they needed more practice due to their unfamiliarity with AR technology. %T13 thought our component-by-component tutorial was good, while T10 suggested that we should have a computer simulation before using the AR device.

\subsubsection{Distraction}
\label{Distraction}
In terms of distraction, participants provided an average rating of 3.03 ($SD=1.62$) on a 7-point Likert scale, where 1 indicates the least distracting and 7 indicates the most distracting. Ten participants thought the system was not distracting but rather enabled them to focus on important visual information. They described VisiMark as ``more useful than distracting'' (T3, T6, T9) and ``a way to eliminate visual noise from a space'' (T5). %As T11 commented, ``I don't think it was distracting, because I knew what I was looking for.'' 
However, two participants (T4, T10) found that the augmentations distracting due to the learning curve.

As for the information intensity, some participants (T2, T6, T8) explicitly mentioned that the current intensity was suitable and not overwhelming. As commented by T8, ``More visuals may be too complex.'' 
However, T10 thought there was too much information in the current system. In the study, VisiMark augmented 0–2 landmarks in one hallway and 4-5 landmarks in one route. However, T10 considered \textit{fewer than three landmarks} in one route with two turns to be more suitable. Additionally, T5, T6, and T14 added that in some visually crowded environments, the augmentations could become overwhelming. As T5 explained, ``Sometimes [augmentations] increased visual noise and didn't particularly help.'' We discuss this issue further in Section \ref{Signboards}, focusing on what types of landmarks should only be augmented in preview.
% \yuhang{you mentioned "some spots", try to specify. I changed to visually crowded, but you can adjust}

Participants (T4, T12, T15-16) also discussed over-reliance issues since they may focus more on augmentations and look around less when wearing the system. %Additionally, two participants (T14, T15) noted that while focusing on the system, they were often distracted by people passing by, trying not to run into them. Reflecting on these attention shifts, four participants () expressed concerns about over-reliance on the system. 
They were concerned that the system offered limited information compared to the entire environment (T15), so that they would only receive information passively, relying on selected landmarks by VisiMark other than picking up landmarks by themselves (T4, T12, T16). 
%and augmentations might overlook some cues even obstacles (T6,T11-12). %As T12 explained, ``I think I did a little better at identifying landmarks without the system.'' 
%Another drawback noted was that receiving information made them less proactive, relying more on selected landmarks than picking up landmarks by themselves (T4,T12,T16).
T11 thus suggested that we should augment more on dangers, especially obstacles on the floor. We discuss this further in Section \ref{Landmarks outside their central view.}.

\subsection{Taxonomy of Landmarks to Augment} 
% \yuhang{this is not taxonomy since i don't see categories as subsections...see Jae's paper as an example of taxonomy: https://arxiv.org/abs/2407.13515}
All participants found the current landmark selection useful because they aligned with the key types of landmarks they commonly used. As explained by T7, ``A lot of the things that you will have projected on the screen here are kind of like what I drew on the mental map or things that I used to navigate.'' %Therefore, %\textbf{\textit{all landmarks that meet PLV's criteria should be augmented}}. 
However, beyond the common landmark categories we derived in the formative study, participants expressed preferences for augmenting additional types of landmarks that they would not be able to identify by themselves in daily life. \colorchange{We report the landmarks to augment in AR below and summarize the taxonomy of landmarks in both real world (from the formative study) and AR contexts in Table \ref{tab:landmark_taxonomies}.}
% The characteristics of landmarks were consistent with those identified in the formative study for PLV. Specifically, for landmark modalities, one participant (T15) noticed tactile landmarks (temperature) in addition to auditory and olfactory landmarks.

\subsubsection{Unique but visually challenging landmarks.} 
Unique but not visually obvious landmarks refer to those that are distinct in the environment for sighted people but difficult for PLV to see, including special objects that are small or lacking sufficient contrast. Participants appreciated that the system highlighted unique objects that might have been overlooked by them due to visual impairments (6/16). For example, T4 commented on a set of green double doors with low contrast augmented by VisiMark, ``For some reason I keep coming back to those [green double doors]. That was very distinct [with augmentations]. Maybe because it's a landmark that I wouldn't take for granted [with my own vision].'' PLV might not have chosen some unique landmarks as reference points without VisiMark, as these landmarks did not stand out visually to them.

%TC:ignore
\begin{table*}[h]
\centering
\small 
\begin{tabular}{>{\raggedright\arraybackslash\color{brownishred}}m{2.8cm} >{\raggedright\arraybackslash\color{brownishred}}m{4.3cm} >{\raggedright\arraybackslash\color{brownishred}}m{5.8cm} >{\raggedright\arraybackslash\color{brownishred}}m{3.4cm}}
\Xhline{2\arrayrulewidth}
\textbf{Contexts} & \textbf{Categories} & \textbf{Subcategories} &\textbf{Examples}\\
\Xhline{2\arrayrulewidth}
\multirow{9}{*}{Real-world Landmarks}& \multirow{3}{*}{Visual Landmarks}&Landmarks with visually salient characteristics \cite{sorrows1999nature} & Large, high contrast objects\\
\cline{3-4}
& &Lighting conditions* & Dark stairs, reflective floors\\
\cline{3-4}
& &Perceived ``silhouette'' of an area* &The hallway as a whole\\
\cline{2-4}
& \multirow{4}{*}{Cognitive Landmarks}& Elements with important functions \cite{raubal2002enriching}&Restaurants, restrooms\\
\cline{3-4}
& &Personal experiences \cite{sorrows1999nature}&Sensitive to numbers\\
\cline{3-4}
& &Danger and emergency* &Danger signs, emergency exits\\
\cline{3-4}
& &Landmarks with accessibility purposes*&Railings, ramps, elevators\\
\cline{2-4}
& \multirow{2}{*}{Structural Landmarks}&Structural elements \cite{sorrows1999nature}&Pillars, doors\\
\cline{3-4}
& &Overall structure of the floor plan*&Size of the space\\
\hline
\multirow{6}{*}{AR Landmarks}& Unique but visually challenging &Unique objects that are small&Small art sculptures\\
\cline{3-4}
&   landmarks& Unique objects with low contrast& Green double doors\\
\cline{2-4}

& Cognitively important but visually challenging landmarks& Recessed or flat cognitive landmarks & Elevators, restrooms\\
\cline{2-4}
& \multirow{2}{*}{Landmarks outside field of view}&Landmarks above eye level &Clocks\\
\cline{3-4}
&&Obstacles below eye level&Floor-level obstacles\\
% \hline
% \multirow{4}{*}{Augmenting Timing}& only in preview&visually obvious
% landmarks\\
% \cline{2-3}
% & \multirow{3}{*}{only in situ}& common yet important facilities\\
% \cline{3-3}
% & &object affordance\\
% \cline{3-3}
% & &small or low-contrast prints\\
\Xhline{2\arrayrulewidth}
\end{tabular}
\caption{\colorchange{Summary of two landmark taxonomies: (1) landmarks commonly used by PLV in real-world navigation derived from the formative study, with * labeling unique subcategories for PLV; and (2) landmarks that PLV preferred to augment in AR.}}
\label{tab:landmark_taxonomies}
\end{table*}
%TC:endignore

\subsubsection{Cognitively important but visually challenging landmarks.}
Visually challenging but cognitively important landmarks are those that hold cognitive significance but lack obvious visual features. They include recessed or flat landmarks, such as elevators and restrooms that are hidden in walls. As T12 noted, ``I need to see the restroom sign clearly to know it's a restroom. And the restrooms look similar to a classroom.'' T2 added, ``[Restrooms] are important, but they're very flat... The icon for the restrooms and the elevators are helpful because they do tend to be recessed or flat so it's easy for me to walk right past them.''

\subsubsection{Landmarks outside field of view.}\label{Landmarks outside their central view.} Landmarks outside PLV's field of view, such as those on walls, floors, or ceilings, could also be difficult to notice and thus being preferred to be augmented by AR. %include those on the wall, floor or ceiling, especially ones that could pose a danger. 
Since PLV, especially people with limited visual field, tend to look at eye level and below, augmenting landmarks outside their central view can reinforce their awareness of these objects (T3, T6, T7, T16). As T16 explained, ``[Augmenting things above eye level] might be nice... I don't look up very much.'' Four participants (T6, T11, T15, T16) expressed a desire to augment obstacles on the floor. Given that PLV are particularly attentive to landmarks related to safety, augmenting floor-level obstacles is important to prevent users from tripping on hazards.

\subsection{Desired Augmentation Designs: Comments and Improvements of VisiMark}
Low vision participants \colorchange{discussed their experiences and provided} insightful feedback for each design element in VisiMark. \colorchange{We report our findings below, and summarize the augmentation experiences and preferences of people with different visual abilities in Table \ref{tab:augmentation_diff_conditions}.}
%Our findings are presented below.
% They also suggested more customization options to further tailor the landmark augmentation system to meet diverse needs. 

\subsubsection{Signboards}
% \yuhang{this and the following sections are too long. Take a look at the length of other subthemes, and use them as a standard... flatten the structure. And remove some less interesting results.}
\label{Signboards}
All participants liked the design of arrows and landmark labels on the Signboard, as they provided an overview of the upcoming hallway structures and landmarks (11/16), enabling self-orientation without additional explorations (8/16). As T5 commented, ``I like knowing what's going to be in the hall before I go there, and trying to find a map in a building [is] usually a pain.'' T3 added, ``[Without the system] it would take me a lot longer to memorize and to understand where I'm at.'' Importantly, participants with depth perception loss found the overview of hallway structures particularly helpful (T5, T11, T14, T15) since it helped them identify the dead ends (T3, T5, T12). And for participants with double vision like T12, they liked the overview of hallway structures because it reduced the need to look around, which would exacerbate their double vision. T12 mentioned that the overview of landmarks was more helpful than in-situ labels for her, as she preferred focusing on visual cues in her central view. %For participants with monocular blindness, the design of Signboards pointing out all possible directions was helpful, as commented by T9, ``It (the system) would help with my legally blind eye to help me with perception and which way to go for sure.''

Participants further offered suggestions to improve Signboard design. T5 and T8 suggested including scales on the Signboards to understand the relationship between the arrows' dimensions and the actual hallways' dimensions. T16 suggested adding small arrows to further connect hallways to enhance understanding of the floor layout. Additionally, T5 and T13 recommended presenting a map of the layout of the whole space in the corner of the Signboard, together with cardinal directions. T13 also hoped for the map to show their real-time location. Overall, participants desired for more comprehensive knowledge of their surroundings, rather than information limited to their nearby vicinity.

\subsubsection{Color-coded Hallways.} Ten participants considered the color-coded hallways as helpful cues for memorization, noting that the hallway colors corresponding to the Signboard helped them confirm they were on the right track (T15-16) and made navigation easier unconsciously (T5, T12, T16). Moreover, participants with depth perception loss especially appreciated the design, as the colored overlay clearly indicated where the hallway ended (T6, T8, T11). As T6 explained, ``I like the [color-coded hallways]. It actually goes a whole length of the hallway.'' However, six participants did not find this design useful, feeling the colored hallways distracting. Nine participants suggested that the color-coded design can be combined with turn-by-turn instructions as a wayfinding support.%a preference for combining the current system with turn-by-turn instructions during the wayfinding task (e.g., always follow the same color-coded floor to the destination). 
%Most participants (15/16) liked the current color selection of color-coded hallways, appreciating the distinct primary colors used. They also suggested additional customization options, including allowing brightness adjustability, transparency adjustability, and more color choices. Specifically, T11 noted that it’s hard for her to distinguish shades, so the system should use contrasting colors. She added that using darker colors as outlines would help meet high contrast demands for PLV.

% However, some participants suggested additional customization options: T2, T4, and T11 recommended allowing brightness adjustability, and T6 suggested transparency adjustability. T7, T8, and T11 mentioned the need for more color choices. T11 noted that it’s hard for her to distinguish shades, so the system should use contrasting colors. She added that using darker colors as outlines would help meet high contrast demands for PLV. In summary, participants expressed a preference for further customization options, suggesting that personalized adjustments in color, brightness, and contrast could enhance the system’s usability for a broader range of low vision needs.

\subsubsection{In-situ Labels}
\label{In-situ labels}

All participants appreciated the in-situ icons and texts, describing them as ``focus points to tie on to'' (T5, T6), which helped them confirm they were on the right track at distance (T7, T9, T11-12). Participants showed different preferences between text and icon labels. Several participants (T4, T11, T15) preferred icons because they were simple, easy to understand, and accessible to people who cannot read. %As explained by T4, ``It's (the icon) simple... but it conveys the same thing.'' T11 added that icons could help people who cannot read. 
In contrast, some participants (T1, T5, T7, T10) preferred text labels for their clarity and details. 
% T5 noted that with the system using five categories of icons instead of a unique icon for each landmark, ``Icons aren't like the most immediately indicative of what it's supposed to be the text is.'' Similarly, T7 suggested, ``I think if they (icons) align more naturally to the environment, like the landmarks (icons) that we're used to seeing and buildings already, that would be nice.''

Since VisiMark used five categories of icons rather than a unique icon for each individual landmark, some participants recommended having unique icons for specific landmarks, such as restrooms (T2, T5, T7, T13, T14) and stairs (T1, T3, T9), to make them more explicit. As T1 explained, stairs were important for PLV due to the risk of falling: ``All I really care about is the stairs.'' Opinions on the number of icon categories varied: while T1 and T6 found the current number of categories (five) sufficient, T7 suggested having no more than 10 categories, and both T5 and T7 preferred a unique icon for each landmark. Offering multiple sets of icon choices, with unique icons for specific landmarks, would be ideal.

As for text clarity, T6 suggested avoiding abbreviations. For instance, instead of using ``AED'', the system should use ``emergency defibrillator.'' Additionally, several participants (T1, T7, T13-14) suggested including more detailed text descriptions, such as specifying gender labels for restrooms. T1 pointed out, ```Green locker is better than `locker.' I think it's just another little piece of information that would help you put it together.'' 

In general, participants preferred clear and informative in-situ labels for both icons and text, suggesting improvements such as unique icons for key landmarks and avoiding abbreviations to enhance clarity and usability.

\begin{table*}[h]
\centering
\small 
\begin{tabular}{>{\raggedright\arraybackslash\color{brownishred}}m{4.5cm} >
% {\raggedright\arraybackslash\color{brownishred}}m{2.7cm} >
% {\raggedright\arraybackslash\color{brownishred}}m{5.8cm} >
{\raggedright\arraybackslash\color{brownishred}}m{12.5cm}}
\Xhline{2\arrayrulewidth}
\textbf{Components} &\textbf{Experiences and Preferences}\\
\Xhline{2\arrayrulewidth}
\multirow{4}{*}{Signboards}& Provide an overview of the upcoming hallway structures and landmarks for all participants\\
\cline{2-2}
&Enable people with depth perception loss to identify dead ends\\
\cline{2-2}
&Reduce active scanning around and facilitate focus for people with double vision\\
\cline{2-2}
&Enable people with field of view loss to notice landmarks in their blind spots (Section \ref{Impact on Landmark Selection})\\
\hline
\multirow{3}{*}{Color-coded Hallways}& Promote route memorization and confirmation for ten participants\\
\cline{2-2}
&Enable people with depth perception loss to better perceive the hallway structure\\
\cline{2-2}
& Distract six participants\\
\hline
\multirow{3}{*}{In-situ Labels}& Facilitate landmark identification and route confirmation at distance for all participants\\
\cline{2-2}
& Seven participants preferred icons as in-situ labels for simplicity and accessibility\\
\cline{2-2}
& Nine participants preferred text as in-situ labels for clarity and details\\
\Xhline{2\arrayrulewidth}
\end{tabular}
\caption{\colorchange{The augmentation experiences and preferences of participants with different visual abilities.}}
\label{tab:augmentation_diff_conditions}
\end{table*}
%TC:endignore

\subsection{Preferences on Augmenting Different Types of Landmarks}
While most landmarks should be augmented both in overview and in situ, we found that participants preferred to augment certain types of landmarks only on the Signboard or only via In-situ Labels.% report what landmarks should be augmented on Signboards vs. what landmarks should be augmented via In-situ Labels below. 

%For Signboard, participants gave opinions on what types of landmarks should only be augmented on Signboards. 
For visually obvious landmarks, while they were visible to low vision participants, participants (T5, T16) still wanted them to be augmented via Signboard to get an overview and avoid the visual distraction caused by In-situ Labels when augmenting already visually salient objects. 
%As T5 explained, ``[Augmenting visually obvious landmarks on the Signboard] makes it really easy [to identify and locate].'' T16 suggested to put visually salient landmarks only on Signboards, which was aligned with 
As T5 commented on a visually obvious window, ``Occasionally [the in-situ augmentation of a window] was a little more visual noise than necessary.'' As such, \textbf{\textit{augmenting visually obvious landmarks only in preview via Signboard}} helps reduce unnecessary visual distractions.

%Participants also noted what types of landmarks are more suitable to be augmented only in situ. 
For facilities with important functions but not unique, such as restrooms, In-situ Labels were the preferred augmentations. Participants (T4, T10, T13, T15) commented that such landmarks were not memorable because they were everywhere. However, these functional facilities played an important role in mental mapping. % as noted by T13, ``So if I just, like in my mind, try to map out windows [and] restrooms, then there could be several windows [and] several restrooms [in augmentations].'' 
Hence, \textbf{\textit{augmenting these common yet important facilities only in situ}} would be more appropriate. 
As T15 indicated, ``It's harder to tell what were restrooms and what weren't. Seeing like the augmented reality, they will point out, what is a restroom, especially if you don't know the layout, it's able to help you.'' Additionally, some participants would like to \textbf{\textit{augment object affordance in situ}}, such as augmenting door handles (T16), edges of the floor or wall (T1), and tread nosing of stairs (T1, T3, T11). 
%and having leading signs between landing in stairs (T11). 
Moreover, \textbf{\textit{small or low-contrast prints were suitable to be enhanced in situ}} (T7, T8, T11), such as signs on doors (T8) or maps (T11). %Other types of landmarks should be augmented both in preview and in situ.

\section{DISCUSSION} 
% \yuhang{The discussion is really rough compared to the other part of the paper. For a paper with 20 pages, the discussion should be at least 2 pages. You have two studies and a system in the paper, but the summary of contribution is just a couple of sentences... need to better highlight the contribution.}

In this paper, we explored how PLV perceive, define, and use landmarks in navigation and mental map construction. We presented VisiMark, which includes Signboards that provide overviews of hallway structures and upcoming landmarks, along with In-situ Labels for individual landmarks. We found that VisiMark received positive feedback in retracing and mental map drawing tasks, shifting PLV's landmark selection from visually obvious objects to more memorable and meaningful ones. We also summarized what PLV prefer to augment in indoor navigation and how to augment them.

In this section, we summarize the two sets of landmark taxonomy for PLV in both real-world and AR contexts. We also derive design implications for future landmark augmentations for PLV and discuss limitations and future directions.

\subsection{Landmark Taxonomies for PLV in Real World and AR Contexts}
One key contribution of this paper is that we derived two sets of landmark taxonomies for PLV in different contexts, including a taxonomy of landmarks that PLV commonly use in real world navigation (i.e., real-world context), and a taxonomy of landmarks that PLV prefer to augment via AR technologies (i.e., AR context). We summarize these taxonomies and highlight the unique landmark selections by PLV compared to prior literature. 

\subsubsection{Landmark Taxonomy for PLV in Real World}
We first derive the landmark taxonomy for PLV by expanding the landmark categories for sighted people \cite{saha2019closing}. %For landmarks PLV commonly used in daily life, as PLV primarily focus on landmarks in visual modality, the way to categorize landmarks for PLV is more similar to for sighted people than for blind people, whose categories emphasize different modalities \cite{saha2019closing}. 
Our study shows that landmarks for PLV align with the visual, cognitive, and structural categories for sighted people, but they include more unique subcategories. For visual landmarks, PLV pay particular attention to changes in lighting conditions and perceived area “silhouettes,” expanding the range of visual landmarks from individual objects to broader spaces. This reflects PLV's needs to perceive the environment holistically, as discerning fine details can be challenging. Similarly, for structural landmarks, PLV show a tendency to know the overall structure of the space, such as the size and layout of the space. For cognitive landmarks, PLV are particularly interested in landmarks that affect their safety and mobility, whether aiding or impeding it, including potential dangers and accessibility facilities. These preferences in landmark selection show PLV's intent to safely understand the surrounding environment from a macro perspective by interpreting information without relying on fine details. % and organize the landmark and wayfinding information into a coherent overall structure.

\subsubsection{Landmark Taxonomy for PLV in AR Augmentations}
Despite the landmark selection strategies developed by PLV, they still faced challenges in landmark perception, thus desire to augment an additional set of landmarks to support navigation. 
%or landmarks PLV would like to augment in AR contexts, we found that PLV are willing to augment all landmarks that meet their criteria. 
Specifically, PLV prefer to augment certain types of landmarks in AR, including unique but not visually obvious landmarks, visually challenging yet cognitively important landmarks, and landmarks outside their field of view, particularly those indicating potential dangers. These preferences arise from the challenges PLV face in identifying and perceiving meaningful landmarks that do not possess salient visual features. In addition to what to augment in AR, we also identified how to augment these landmarks. To eliminate visual noise, visually obvious landmarks should only be augmented in preview, while common yet important facilities, object affordances, and small or low-contrast prints are more preferred to be augmented only in situ. The other types of landmarks should be augmented both in preview and in situ. In general, PLV prefer to enhance landmarks that are difficult to identify independently, while maintaining an appropriate amount of visual stimuli in AR.

\subsection{Design Implications for Landmark Augmentations}
Throughout the study, low vision participants shared their preferred augmentation designs for future landmark augmentation systems. In this section, we summarize and expand on these design implications.

\textbf{\textit{Prioritizing Safety.}} During our experiment, low vision participants kept emphasizing the importance of safety, suggesting the inclusion of obstacle warnings on the floor and paying particular attention to landmarks with safety and accessibility implications. Since the AR headset captures their primary focus, highlighting potential dangers becomes crucial to prevent accidents caused by attention shifts. Future research should thus explore how to design effective safety warnings on AR-based navigation systems. \colorchange{One approach could be incorporating multi-modal feedback \cite{zhao2020effectiveness,bai2017smart}, such as generating audio alerts for obstacles. Additionally, integrating eye-tracking technology to design augmentations for low vision users offers another promising solution. This approach has been explored to enhance reading experiences \cite{wang2023understanding, wang2024gazeprompt}. Similar ideas could to applied to display safety warnings based on user's gaze, especially when critical hazards are overlooked.}

\textbf{\textit{Augmenting Object Affordances.}} Beyond the entire landmark, participants also expressed a need for augmenting the landmark affordance, such as door handles and stair tread nosing (see Section \ref{In-situ labels}), enabling them to easily perceive important details of these landmarks and better use these facilities. While current computer vision technology excels in whole-object recognition \cite{he2016deep,redmon2016you,liu2021swin}, affordance recognition models are still in its infancy \cite{chuang2018learning,luddecke2017learning, luo2022learning}. \colorchange{A recent work by Lee et al. \cite{lee2024cookar} has developed a kitchen tool affordance dataset and fine-tuned an object segmentation model to distinguish different interactive components of kitchen tools (e.g., knife blade vs. handle) for low vision users. Future research should expand the scope from the kitchen scenario to broader daily activities (e.g., navigation), identifying critical object affordance for low vision and developing effective affordance datasets and recognition models.}

% and   affordance detection and augmentations have already been explored in specific contexts, such as cooking scenarios through systems like CookAR \cite{lee2024cookar}. Expanding this research to other daily tasks, such as navigation or shopping, represents a promising research direction.} \yuhang{cite CookAR and discuss how we should expand}

\textbf{\textit{Understanding Surroundings Ahead.}} Despite possible visual limitations, PLV showed a tendency to know the structure of the space ahead. This is also reflected in their suggestions that Signboards should offer a broader understanding of their surroundings, beyond just their nearby environment (see Section \ref{Signboards}). This aligns with the process of cognitive map construction, which progresses from landmarks to routes and ultimately forms a comprehensive cognitive map \cite{siegel1975development}. Future landmark augmentation systems should not only enhance overviews of the nearby environment and local landmarks, but also offer a more comprehensive overview of the larger surroundings. As computer vision technology can only recognize the immediate vicinity, researchers should \colorchange{consider crowd-sourcing or community-based method \cite{howe2006rise,brambilla2014community} to achieve broader landmark information, enabling various stakeholders (e.g., friends, family, or local community) to label and share important landmarks. Such community-based method has been successfully used and deployed in accessibility research, such as Project Sidewalk \cite{saha2019project}, and could potentially be expanded to landmark labeling and augmentations for low vision users}. 

%The system could then use these labeled landmarks to generate a more comprehensive overview of the space ahead.

\colorchange{
\textbf{\textit{Gaining Agency in Landmark Selection.}} While appreciating VisiMark's assistance in landmark perception, PLV desired more control over landmark selection. They raised concerns about passively receiving landmark information selected by VisiMark and potential information overload if all types of landmarks are augmented. Future landmark augmentation systems should offer sufficient agency to low vision users, allowing them to adjust the types of landmarks to display as well as labeling additional personally relevant landmarks to facilitate memory and mental map development \cite{nuhn2017personal}. To achieve this goal, researchers should consider how to design an accessible landmark selection and labeling interface for low vision users given their challenges of recognizing certain landmarks. For example, such systems may incorporate general visual augmentation methods (e.g., magnification, edge enhancement) \cite{zhao2015foresee,stearns2018design} to enhance the environment's overall visibility, or integrate object recognition to help users identify objects of interest as potential landmarks \cite{liu2019edge, ghasemi2022deep}. Interaction techniques in AR should also be designed to enable low vision users to easily select and label landmarks. Prior research by Zhao et al. \cite{zhao2019designing_interaction} has investigated accessible interaction techniques via speech and touch gestures on smartwatch for low vision users to control different visual augmentations on AR glasses. Future research could consider expanding these techniques to better support landmark selection and labeling. }

\subsection{Limitations and Future Directions}
Our study has some limitations in both system design and user studies. From the system design and implementation perspective, we reveal below limitations and suggest potential future directions:

\textbf{\textit{Integrating AI into AR.}} Our current implementation relies on a pre-scanned environment to reduce the impact of technical limitations (e.g., recognition errors) on user experience. To enhance real-world applicability, we should implement AI-powered AR for real-time landmark recognition and augmentation in the future. 

\colorchange{\textbf{\textit{Adapting to more commonly-available AR platforms.}} Current VisiMark is built on a head-mounted AR device to enable hands-free experience and mitigate attention switching between the real world and an additional display (e.g., if using mobile AR) \cite{thomas2009wearable}. However, some PLV found the form factor of head-mounted AR uncomfortable. Future research should consider adapting VisiMark to more commonly-used and accepted devices, such as smartphone or smartwatches. Importantly, different feedback modalities and designs should be considered for different platforms. For example, as opposed to visual augmentations on head-mounted AR, vibrations on the smartwatch or auditory descriptions from the smartphone could be used to avoid constant visual attention shifts between the device and the real-world environment. }

\colorchange{\textbf{\textit{Allowing more Flexible Customization.}} Current VisiMark allows users to adjust the height, size, and colors of the augmentations. However, more flexible customization should be considered due to low vision users' diverse visual abilities. For example, since VisiMark anchors Signboards and In-situ Labels to certain physical locations, participants with visual field loss may have to look around to view the whole augmentations. Future design should introduce more customization options, such as repositioning augmentations (or different components of an augmentation, e.g., arrows on the Signboard) to user’s functional visual areas rather than relying solely on fixed physical positions, enabling users with diverse visual conditions to better access and fully utilize the system.} %\yuhang{add another limitation that we should consider more customization options, e.g., repositioning different components of the signboard for people with field of view loss. mentioned by 2AC.}

\colorchange{From the user study perspective}, first, our experiments were conducted indoors due to the current visibility constraints of head-mounted AR feedback in outdoor settings. Outdoor environments are more dynamic and more visually complex. Future work should explore additional landmark augmentation designs and preferences for PLV in outdoor environments. 
\colorchange{Second, during our evaluation, we provided verbal turn-by-turn instructions to guide the participants in the initial exploration phase. This decision is informed by low vision people's daily experience as they commonly ask for verbal instructions from others in navigation \cite{szpiro2016finding}. However, reliance on verbal instructions rather than visual landmarks may influence our measurements---especially retracing and mental map of the route segments---potentially reducing the observed differences in performance between VisiMark and the baseline. In fact, one participant (T10) in our study mentioned relying more on the verbal instructions to memorize the route. %Additionally, delivering instructions via the auditory channel, rather than visually, introduced a different input modality that could have affected participants’ route memorization. 
%This may have impacted metrics such as Correctness of Turns, Total Landmark Recall, Retracing Time, and Retracing Accuracy. 
Future research could consider other instruction alternatives, such as visual indicators or landmark-based instructions, to triangulate the results.} 
\colorchange{Third, we had a relatively small number of participants (six) in our formative study to provide design implications. Future research should include a larger group of participants with a broader range of visual conditions to enrich the data.} Lastly, our current study employed relatively \colorchange{short and simple routes, which may also lead to insignificant difference in performance and restrict the assessment of long-term mental map development. Future evaluation should consider more complex environments to better simulate real world conditions and assess the impact and practicality of VisiMark.}

\section{CONCLUSION}
We contributed to the first exploration of how people with low vision (PLV) perceive, define, and use landmarks in navigation through a contextual inquiry study. We found that the categories and usage of landmarks by PLV were generally similar to those by sighted people, but included additional subcategories. For example, PLV paid more attention to ``silhouette'' of an area and landmarks with safety and accessibility implications. We designed VisiMark, a landmark augmentation AR interface. Through a user study with 16 low vision participants, we evaluated the system and found that VisiMark received positive feedback from PLV both in retracing and mental map building tasks. Additionally, VisiMark enabled PLV to better identify landmarks they prefer but could not easily perceive before, and changed PLV's landmark selection from only visually-salient objects to cognitive landmarks that are more unique and important. Our work explores the design space for future AR landmark augmentations for PLV.

\begin{acks}
Research reported in this publication was supported by the National Eye Institute of the National Institutes of Health under Award Number R01EY037100. The content is solely the responsibility of the authors and does not necessarily represent the official views of the National Institutes of Health.
\end{acks}

\bibliographystyle{ACM-Reference-Format}
\bibliography{sections/references}

\appendix
%TC:ignore
\section{Theme Table}
\label{Codebook for Evaluation}
\begin{table*}[ht]
\scriptsize
\centering
\begin{tabular}
% {>{\centering\arraybackslash}p{3cm}>{\centering\arraybackslash}p{3cm}>{\centering\arraybackslash}p{3cm}}
{p{3.5cm} p{4.8cm} p{8.5cm}}
\toprule
  \textbf{Themes} & \textbf{Sub-themes} &  \textbf{Codes}\\
% \Xhline{2\arrayrulewidth}
\hline
Effect of Landmark Augmentations on Route Retracing & \multirow{2}{*}{Effective retracing} & \multirow{1}{*}{navigate easier; memorize the turns}\\
\hline
\multirow{6}{*}{Effect of Landmark Augmentations on} & Perceive hallway structures & notice hallway structures more than before; memorize hallway lengths\\
\cline{2-3}
\multirow{6}{*}{Mental Map Development} & \multirow{2}{*}{Increased Focus on Landmarks} & notice landmarks; identify landmarks; memorize landmarks; locate landmarks; increased reliance on landmarks in navigation\\
\cline{2-3}
& \multirow{4}{*}{Shift in Landmark Selection} & help notice things may be overlooked; monocular blindness: things in blind spot; pick out important landmarks better; help find functional facilities; offer more information to memorize; start paying attention to once augmented landmarks even without wearing the system; conflict with their own way of identifying landmarks\\
\hline
\multirow{6}{*}{Experiences with VisiMark} & Effectiveness&  effective in retracing; effective in mental map building\\
\cline{2-3}
& \multirow{2}{*}{Comfort}& mentally comfortable; more comfortable without the system; device uncomfort; uncomfortable in public\\
\cline{2-3}
& Learnability& easy to use and understand; short learning curve; tutorial\\
\cline{2-3}
& \multirow{2}{*}{Distraction}& more useful than distracting; eliminate visual noise from a space; very distracting because of learning curve; not overwhelming; overwhelming at some spots; limit the number of augmented landmarks\\
\hline
\multirow{4}{*}{Taxonomy of Landmarks to Augment} & Current landmarks in VisiMark & similar to those used in wayfinding and mental maps\\
\cline{2-3}
& Unique but not visually obvious landmarks& green double doors\\
\cline{2-3}
& Visually challenging but cognitively important landmarks & recessed or flat landmarks; elevators; restrooms\\
\cline{2-3}
& Landmarks outside their central view, especially dangers & landmarks above eye level; obstacles on the floor\\
\hline
\multirow{2}{*}{When the Augmentations Should Occur} & What to augment only only in preview & visually salient landmarks\\
\cline{2-3}
& What to augment only in situ & common yet important facilities; affordance; small or low contrast prints\\
\hline
\multirow{14}{*}{Desired Augmentation Designs} & \multirow{7}{*}{Signboards} &  have an overview ahead; locate oneself without extra trips; depth perception issues: providing hallway lengths; depth perception issues: help identify dead end; double vision: prefer signboards in central view; monocular blindness: point out possible directions; have scales; small arrows of further connecting hallways; maps of the general layout; colors are helpful cues to remember; confirm on the right track; easier navigation unconsciously; not turn-by-turn color-coded hallways distracting; distinct current colors; primary colors; allow brightness adjustability; allow transparency adjustability; allow more color choices; add dark colored outlines\\
\cline{2-3}
& \multirow{3}{*}{In-situ labels}& focus points to tie on; confirm on the right track from a distance; icons are simpler but convey same information; icons help people who cannot read; number of icon categories; unique icon categories; texts are more indicative; should not use abbreviation; more details in descriptions\\
\cline{2-3}
& \multirow{2}{*}{Further customization options} & specialize based on the building environment; add ability to turn on and off some components; customize personal layers; add ability to zoom in; add ability to adjust position of augmentations\\
\bottomrule
\end{tabular}
\caption{Themes and Codebook.}
\label{tab:Themes and Codebook}
\end{table*}

\section{Interview Questions for the Formative Study}
\label{Interview Questions for the Formative Study}
\subsection{Initial Interview Questions}
\begin{enumerate}
    \item What is your name?
    \item What is your age?
    \item What gender do you identify with?
    \item What is your visual condition?
    \item Are you considered Legally blind?
\begin{enumerate}
    \item What is your diagnosis? 
    \item What is your visual acuity?
    \item What is your field of view?
    \item What is your contrast sensitivity?
    \item What is your color vision?
    \item What is your light sensitivity? 
    \item What is your eepth perception? 
\end{enumerate}
    \item How long have you had this visual condition?
\begin{enumerate}
    \item Is this condition progressive or stable?
\end{enumerate}
    \item How do you usually complete a navigation task?
    \item Do you use any technology to navigate regularly outdoors and indoors?
\begin{enumerate}
    \item If yes, what technology do you use?
\end{enumerate}
    \item Do you pay attention to any landmarks during navigation? What landmarks?
    \item Do you use any technology to help you perceive the landmarks?
    \item Will your choice of landmarks change due to familiarity?
    \item Do you have any prior experience with Augmented Reality (Google glasses, phone application, HoloLens)? 
\begin{enumerate}
    \item If yes, could you please share the experience?
\end{enumerate}
    \item Are you currently familiar with the campus building?
\end{enumerate}

\subsection{Exit Interview Questions}
\begin{enumerate}
\item How do you determine landmarks? Do you prioritize certain landmark features (e.g., color, size, shape)?
\begin{enumerate}
\item Why do you pay attention to a specific landmark during the navigation task?
\end{enumerate}
\item What type of landmark modalities do you prefer?
\item How do you use your landmarks during navigation (including wayfinding and mental map building)?
\item Do you look for landmarks the first time you visit a place, or only after you’ve been there a few times?
\item Will your choice of landmarks change due to familiarity?
\item How important do you consider landmarks for indoor wayfinding? Could you please offer a score between 1 and 5, with 1 stands for least important and 5 means most important?
\item How important do you consider landmarks for developing a mental map indoors? Could you please offer a score between 1 and 5, with 1 stands for least important and 5 means most important?
\item What types of landmarks would you like to see augmented? Are there any specific landmarks you would prefer to use but find challenging due to visual limitations?
\item What kind of landmark augmentations you would like to have? (e.g., enlarging, outlining, etc.)
\item What modalities of landmark augmentations you would like to have? (e.g., visual, audio, haptic, etc.)
\end{enumerate}

\section{Interview Questions for the Final Evaluation}
\label{Interview Questions for the Evaluation}
\subsection{Initial Interview Questions}
\begin{enumerate}
    \item What is your name?
    \item What is your age?
    \item What gender do you identify with?
    \item What is your visual condition?
    \item Are you considered Legally blind?
\begin{enumerate}
    \item What is your diagnosis? 
    \item What is your visual acuity?
    \item What is your field of view?
    \item What is your contrast sensitivity?
    \item What is your color vision?
    \item What is your light sensitivity? 
    \item What is your eepth perception? 
\end{enumerate}
    \item How long have you had this visual condition?
\begin{enumerate}
    \item Is this condition progressive or stable?
\end{enumerate}
    \item Do you use any technology to navigate regularly outdoors and indoors?
\begin{enumerate}
    \item If yes, what technology do you use?
\end{enumerate}
    \item Do you pay attention to any landmarks during navigation? What landmarks?
    \item Do you use any technology to help you perceive the landmarks?
    \item Do you have any prior experience with Augmented Reality (Google glasses, phone application, HoloLens)? 
\begin{enumerate}
    \item If yes, could you please share the experience?
\end{enumerate}
    \item Are you currently familiar with navigating inside this building?
\end{enumerate}

\subsection{Exit Interview Questions}
\begin{enumerate}
 \item Let’s first talk about your landmark choices in the four trials. (Based on the mental map) Why do you pick this specific landmark?

 \item Our systems select and augment certain types of landmarks for you. Do you think they are useful or not? Why? What landmarks do you prefer to be augmented in indoor navigation? 

\item For [each element], how do you like it? How does this design affect your understanding of the route? How do you want to improve it?
\begin{enumerate}
\item The presentation of the structure of the hallways (e.g., direction, width, length, color-coding, the presentation of deadends)
\item The icons and texts of the landmarks on the signboard.
\item In-situ elements.
\end{enumerate}

\item Effectiveness: How effective do you think of the system? Could you please offer a score between 1 and 7, with 1 stands for least effective and 7 means most effective?
\item Comfortable: How comfortable do you think of the system? Could you please offer a score between 1 and 7, with 1 stands for least comfortable and 7 means most comfortable?

\item Distraction/Load: How distracting do you think of the system? Could you please offer a score between 1 and 7, with 1 stands for least distracting and 7 means most distracting ?
\item Learnability: How easy to understand or learn do you think of the system? Could you please offer a score between 1 and 7, with 1 stands for least easy to use and 7 means most easy to use?

\item Any ideas for other designs of augmentations to support your indoor navigation and mental model development?

\end{enumerate}

%TC:endignore

\end{document}